\documentclass[english,useAMS,usenatbib, letterpaper]{mn2e}
\pdfoutput=1
\usepackage[T1]{fontenc}
\usepackage[latin1]{inputenc}
\usepackage{graphicx}
\usepackage{amssymb}
\usepackage{array}
\usepackage{textcomp}
\usepackage{graphicx}
\usepackage{amssymb}
\usepackage{array}
\providecommand{\tabularnewline}{\\}
\usepackage{booktabs}
\usepackage{babel}
\usepackage{times}
\usepackage{url}
\usepackage{threeparttablex}
\usepackage{subfigure}
\title[ Multi-Wavelength Modelling of the $\beta$ Leo Debris Disc]{  Multi-Wavelength Modelling of the $\beta$ Leo Debris Disc: 1, 2 or 3 planetesimal populations? \thanks{\emph{Herschel is an ESA space observatory with science instruments provided by European-led Principal Investigator consortia and with important participation from NASA.}} }

 \author [Churcher et al.]{L J Churcher  $^{1}$\thanks{e-mail:ljc51@ast.cam.ac.uk},  M.C.~Wyatt$^{1}$, G. Duch\^{e}ne$^{2,3}$,      
              B. Sibthorpe$^{4}$,	  G. Kennedy$^{1}$,
 \newauthor	  B. C. Matthews$^{5,6}$,  P. Kalas$^{3}$, J. Greaves$^{7}$, K. Su $^{8}$ and G. Rieke$^{8}$\\$^1$Institue of Astronomy, University of Cambridge, Madingley Road, Cambridge, UK, CB3 0HA\\ $^2$ Department of Astronomy, University of California, 601 Campbell Hall, Berkeley, CA, U.S.A., 94720\\$^3$ Laboratoire d'Astrophysique, Observatoire de Grenoble, Universit\'{e} J. Fourier, CNRS, France\\$^4$ UK Astronomy Technology Center, Royal Observatory, Blackford Hill, Edinburgh EH9 3HJ, UK\\$^5$ Herzberg Institute of Astrophysics, National Research Council Canada, 5071 West Saanich Road., Victoria, BC, Canada, V9E 2E7\\$^6$ University of Victoria, Finnerty Road, Victoria, BC, V8W 3P6 Canada\\$^7$ School of Physics and Astronomy, University of St Andrews,  North Haugh, St Andrews, Fife KY16 9SS, UK\\$^8$Steward Observatory University of Arizona 933 N Cherry Avenue Tucson, AZ 85721}

\begin{document}
\maketitle
\date{Accepted:  Submitted: }

\pagerange{\pageref{firstpage}--\pageref{lastpage}} \pubyear{2010}
\label{firstpage} 
\begin{abstract}

In this paper we present a model of the $\beta$ Leo debris disc, with an emphasis on modelling the resolved PACS images obtained as part of the Herschel key programme DEBRIS. We also present new SPIRE images of the disc at 250 $\micron$, as well as new constraints on the disc from SCUBA-2, mid-IR and scattered light imaging. Combining all available observational constraints, we find three possible models for the $\beta$ Leo (HD102647) debris disc: (i) A 2 component model, comprised of a hot component at 2 AU  and a cold component from 15-70 AU. (ii) A 3 component model with hot dust at 2 AU, warm dust at 9 AU, and a cold component from 30-70 AU, is equally valid since the cold emission is not resolved within 30 AU. (iii)  A somewhat less likely possibility is that the system consists of a single very eccentric planetesimal population, with pericentres at 2 AU and apocentres at 65 AU. Thus, despite the wealth of observational constraints significant ambiguities remain; deep mid-IR and scattered light imaging of the dust distribution within 30 AU seems the most promising method to resolve the degeneracy. We discuss the implications for the possible planetary system architecture; e.g., the 2 component model suggests planets may exist at 2-15 AU, while the 3 component model suggests planets between 2-30 AU with a stable region containing the dust belt at 9 AU, and there should be no planets between 2-65 AU in the eccentric planetesimal model. We suggest that the hot dust may originate in the disintegration of comets scattered in from the cold disc, and examine all A stars known to harbour both hot and cold dust to consider the possibility that the ratio of hot and cold dust luminosities is indicative of the intervening planetary system architecture.

\end{abstract}

\section{Introduction}

Debris discs are distributions of dust and planetesimals with radii of 1-1000 AU around main
sequence stars (See \citealp{2008ARAA..46..339W}
for a recent review). The dust grains in these discs are small, and so cannot
be primordial as they would have been blown out of the system by radiation
pressure on timescales shorter than the stellar age. The
dust must be continually  replenished from a population of colliding
planetesimals, thought to contain bodies up to $\sim$ 1km
in size (\citealp{2002MNRAS.334..589W}). However, as debris discs age
the planetesimal population is ground down, so discs become less massive
and fainter (\citealp{2003ApJ...598..626D}). At far-infrared and sub-millimetre
wavelengths debris discs are optically thin, the disc to star contrast
is favourable, and these wavelengths are sensitive to the large (up
to $\sim$1mm) grains that dominate the dust mass in debris
discs. 

The dust morphology of a debris disc can be shaped by planets
in the system so resolved images of discs help constrain models of structure
and evolution of planetary systems. Resolved images can indicate that infrared excess is being produced by multiple dust populations and can also break the
degeneracy between the radial location of the dust and its temperature.

The DEBRIS (Disc Emission via a Bias-free Reconnaissance in the Infrared/Sub-mm)
survey (Matthews et al. in Prep, \citealp{ 2010MNRAS.403.1089P}), is an Open Time Key Program on the  \emph{Herschel Space Observatory} which uses PACS (Photodetector Array Camera and Spectrometer \citealp{2010AA...518L...2P}) and SPIRE (Spectral and Photometric
Imaging REceiver, \citealp{2010AA...518L...3G}) to detect, resolve and characterise debris discs
around a volume-limited sample of 446 A through M stars (\citealp{2010AA...518L.135M}).
$\beta$ Leo (HD 102647) was observed at 100 $\micron$ and 160 $\micron$ with
PACS as part of the DEBRIS survey Science Demonstration Phase (\citealp{2010AA...518L.135M}).

$\beta$ Leo (A3V, $L_{*}$=14.0L$_{\odot}$) is a $\delta$ Scuti type star
at a distance of 11.1 pc. The infrared excess around this main-sequence
star was first discovered by \emph{IRAS} (Infrared Astronomical Satellite \citealp{1992AAS...96..625O}),
then confirmed with \emph{Spitzer} (\citealp{2006ApJ...653..675S}). The excess
was unresolved in mid-IR imaging (\citealp{2001AJ....122.2047J,2009ApJ...691.1896A},
$\S$ 2.2) although differences between the \emph{IRAS}  and \emph{ISO} (Infrared Space
Observatory) fluxes led \citet{2002AA...387..285L} to suggest
that the disc emission may be somewhat extended in the ISO beam (52''
aperture). \citet{2006ApJS..166..351C} obtained a Spitzer IRS (Spitzer Infrared Spectrograph) spectrum of
$\beta$ Leo (See $\S$ 3.2) and found a featureless continuum spectrum
consistent with dust at $\sim$120 K located at 19 AU from
the star. A very hot excess has also been partially resolved using
infrared interferometry with the FLUOR (Fiber Linked Unit for Recombination) instrument at the CHARA (Center for High Angular Resolution Astronomy) array at 2 $\micron$ and BLINC (Bracewell Infrared Nulling Cryostat) at 10 $\micron$ (\citealp{2009ApJ...691.1896A}, Stock et al. (2010), see $\S$ 2.8 for more details). 

$\beta$ Leo is thought to be a member of the IC2391 moving group, giving an age of 45 Myr (\citealp{2010AJ....140..713N}). The age of this source determined from
isochrone fitting  is  50-331 Myr  (\citealp{1999AA...348..897L,2001ApJ...546..352S}). \citet{2004AA...426..601D} derive an age from measuring the stellar radius and suggest 100 Myr. We assume an age of 45 Myr for this paper. $\beta$ Leo does not have any known
companions within a few arcsecs of the star. The Washington Double
Star (WDS) catalogue lists three companions with common proper motion for $\beta$ Leo. These stars are located
from 40$\arcsec$ to 240$\arcsec$ from the primary with V magnitude differences
of 6.3 to 13 (\citealp{1997A&AS..125..523W}); these stars are not, however, physically associated with $\beta$ Leo (\citealp{2010MNRAS.403.1089P}). $\beta$ Leo was also included in a high precision radial velocity survey
of early type dwarfs with HARPS (\citealp{2009AA...495..335L}) for
planetary or brown dwarf mass candidates and was found to have no
companions with mass >4.2M$_{jup}$ with periods <10 days with 99.7 percent
probability.

This paper uses multi-wavelength modelling of the $\beta$ Leo debris
disc including PACS and SPIRE observations to present a self-consistent explanation
of the system. In \S2 we present the observations, including 100 and 160 $\micron$ PACS imaging
(previously published in \citealp{2010AA...518L.135M}) and archive
Gemini MICHELLE 12 $\micron$ and 18 $\micron$ imaging. In \S3 we confront the observations with
models to determine the disc parameters. In \S4 we discuss the implication
of the inferred structure for the status of planet formation in this 
system.

\section{Observations}

\subsection{Fitting the Stellar Photosphere}

Being the 5th nearest A-type star, $\beta$ Leo is bright and saturated in modern
surveys such as 2MASS. In addition, the infrared excess has a hot component, which
contributes to bands typically used for modelling stellar photospheres
(\citealp{2009ApJ...691.1896A, 2010ApJ...724.1238S}). We use the equivalent 2MASS $K_s$ magnitude of 1.93 from \citet{2010ApJ...724.1238S}, and include mean UBV
and \emph{Hipparcos} $H_p$ photometry
(\citealp{2006yCat.2168....0M,1997yCat.1239....0E}). The best fitting
\citet{2003IAUS..210P.A20C} model, found by a $\chi^2$ minimisation method, has $T_{\rm
eff} = 8660$K, $L_\star = 14 L_\odot$, and $R_\star = 1.66 R_\odot$. In the MIPS 24 $\micron$ band we predict $1171 \pm 15$ mJy, 1\% lower, but consistent with the value in  \citet{2010ApJ...724.1238S}. The predicted stellar flux densities
at the PACS effective wavelengths of 100 and 160 $\mu$m are $64 \pm 1$ and $26 \pm 0.5$
mJy respectively. These are 9\% higher than the values presented in Matthews et al. (2010) due to improvements in the fitting procedure and data used.

\subsection{Herschel PACS Observations}

Observations of the $\beta$ Leo disc at 100 $\micron$ and 160 $\micron$ were taken
with the ESA \emph{Herschel Space Observatory} (\citealp{2010AA...518L...1P})
using the PACS (\citealp{2010AA...518L...2P}) instrument in photometry
mode as part of the Science Demonstration Phase observations for the DEBRIS survey.
DEBRIS is a flux-limited survey which observes each target to a uniform
depth of 1.5 mJy/beam at 100 $\micron$. The images were first presented
in \citealp{2010AA...518L.135M}. $\beta$ Leo was observed in small scan-map
mode (see PACS Observers' Manual\footnote{PACS operating Manual:\url{ http://herschel.esac.esa.int/Docs/PACS/html/pacs_om.html}} for details). Scan map observations
had eight repeats in a single scan direction at a scan rate of 20''/s.
Four 3' scan legs were performed per map with a 2'' separation between
legs. The total observing time was 1220s.

These data were reduced using the Herschel Interactive Processing
Environment (HIPE \citet{Ott2010}). Maps were obtained via the default
PACS naive map-making method photProject in HIPE. The data were pre-filtered
to remove low-frequency (1/f) noise using a boxcar filter with a width of
98''. All bright sources in the map were masked prior to filtering
to avoid ringing type artifacts. 

\begin{figure*} \label{PACS}
\caption{Images of the 100 (left) and 160 $\micron$ (right) emission for $\beta$
Leo taken with PACS. The images have had a PSF scaled to the
stellar flux (64 mJy and 26 mJy at 100 and 160 $\micron$) subtracted, and
hence are maps of the excess emission. The pixel scale in 1'' per
pixel at 100 $\micron$ and 2\arcsec per pixel at 160 $\micron$.\label{PACS}}

\includegraphics[width=7cm,]{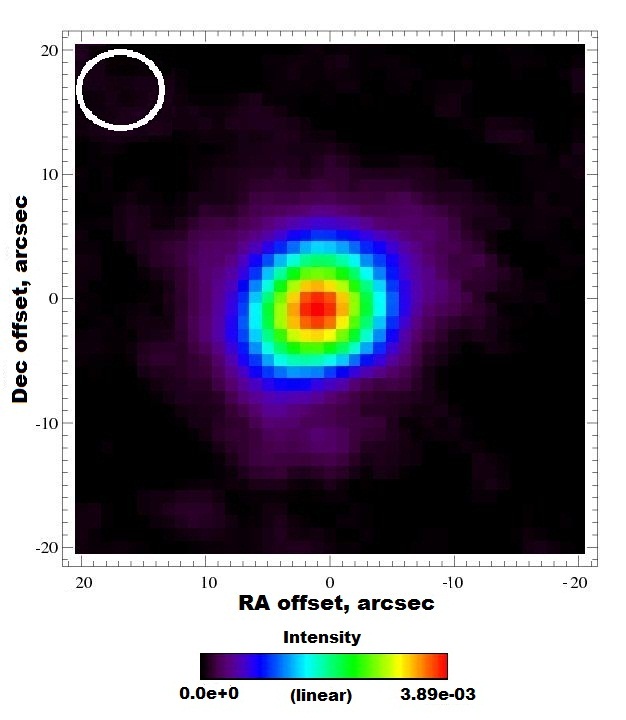}\includegraphics[width=7.5cm]{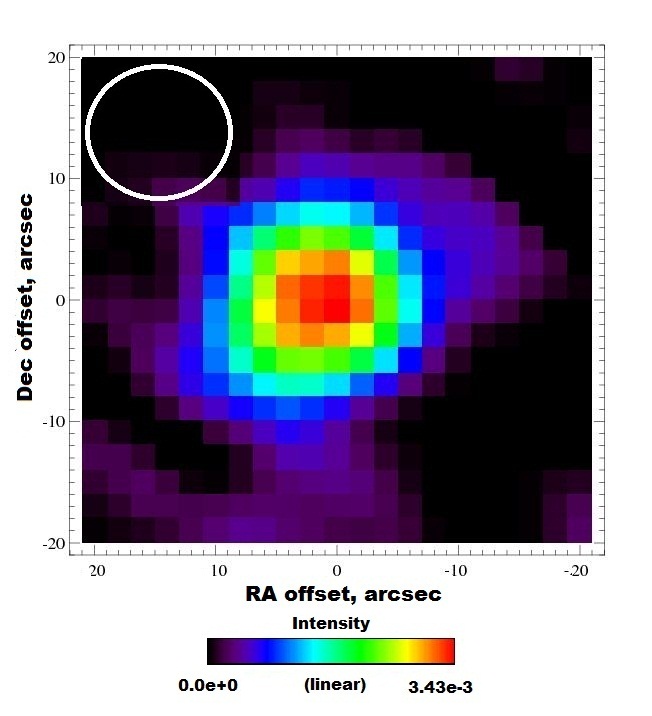}
\end{figure*}

Photometry on the maps of $\beta$ Leo (Figure \ref{PACS}) using a 13'' radius circular aperture centred on the emission peaks yielded 
fluxes of 480$\pm$30 mJy at 100 $\micron$ and 215$\pm$32 mJy at 160 $\mu$m, respectively.
These fluxes greatly exceed the rms noise levels in the maps of 1.4 mJy/beam at 100 $\micron$ and
3.1 mJy/beam at 160 $\micron$. These values have been colour corrected  using the values from the PACS Observers Manual  assuming a temperature of 120K corresponding to the temperature of the disc estimated from a blackbody fit to the SED. The quoted errors on these fluxes do not include calibration uncertainties
which are estimated to be 10 and 20\% at 100 and 160 $\micron$ (\citealp{2010AA...518L...2P}). These were
combined in quadrature with statistical uncertainties from the rms
levels in the maps. The stellar flux for fitting the photosphere  is 64 mJy and 26 mJy at 100 and
160 $\micron$, respectively. This gives excess fluxes of 416$\pm$30 mJy
at 100 $\micron$ and 189$\pm$32 mJy at 160 $\micron$. These values differ  from those in \citealp{2010AA...518L.135M}, which quoted  500$\pm$50 mJy at 100 $\micron$ and 230$\pm$48 mJy at 160 $\micron$ in a 20\arcsec radius aperture. This difference is because the data presented in this paper were rereduced using a different filter scale and a more recent version of HIPE (4.2.0 developer's build) and a smaller aperture was used.

The maps of $\beta$
Leo (Figure \ref{PACS}) appear extended compared to the beam (indicated
by the circle in the top left corner of the images). At 100 $\micron$
the nominal beam size is $6.^{\prime\prime}6\times6.^{\prime\prime}9$. At 160 $\micron$ it
is $10.^{\prime\prime}7\times12.^{\prime\prime}1$. We have obtained 2 bright point source
images (Vesta and $\alpha$ Boo) taken in the same observing mode
as the $\beta$ Leo data to serve as PSF references. These have beams sizes
of $6.^{\prime\prime}6\times6.^{\prime\prime}9$ at 100 $\micron$ and $10.^{\prime\prime}5\times11.5^{\prime\prime}$
at 160 $\micron$ for Vesta and $6.^{\prime\prime}5\times6.^{\prime\prime}8$ at 100 $\micron$
and $10.^{\prime\prime}2\times11.^{\prime\prime}7$ at 160 $\micron$ for $\alpha$ Boo.
 Fitting a Gaussian to the $\beta$ Leo image gives a FWHM at 100 $\micron$ of 9.2\arcsec$\pm$0.1$\times$10.4\arcsec$\pm$0.1. At 160 $\micron$ the FWHM of a Gaussian fitted to the image is 13.6$^{\prime\prime}$$\pm$0.2$\times$12.0$^{\prime\prime}$$\pm$0.2. Figure \ref{PACS} show the $\beta$ Leo images after subtraction of the $\alpha$ Boo
PSF scaled to the appropriate stellar flux. The PSF has an asymmetric
three lobed structure and was rotated to the same spacecraft angle
at which the $\beta$ Leo data were taken at to ensure accurate subtraction.

$\beta$ Leo appears extended in all directions with respect to the PSF,
with the major axis of the disc at a position angle of 125$^{\circ}$$\pm$15$^{\circ}$
E of N, which is the mean of the position angles of the Gaussians fitted at 100 (PA: 118$^{\circ}$) and 160 $\micron$ (PA: 132.5$^{\circ}$). The ratios of the semi-minor to semi-major axis of the Gaussians (0.84 at 100 $\micron$ and  0.88 at 160$\micron$) give a mean inclination of 57$\pm7^{\circ}$ from edge on. The stellar subtraction was also done using the Vesta PSF to examine the effect of PSF variation on the residuals, and resulted in no significant change in the width of a Gaussian fitted to the disc, and no significant change in the flux in a 13\arcsec radius aperture.

\subsection{MICHELLE Mid-Infrared Observations}

Mid-IR observations of $\beta$ Leo taken with MICHELLE on Gemini North for $\beta$ Leo
were retrieved from the Gemini Science Archive (Program Names: GN-2007A-C-10 [PI: Beichman]
and GN-2006A-Q-10 [PI: Moerchen]). These data are detailed in Table 1 and were taken
with filters Qa ($\lambda_{c}$=18.1 $\micron $, $\Delta\lambda$=1.51 $\micron $)
and N' ($\lambda_{c}$=11.3 $\micron$, $\Delta\lambda$=1.07 $\micron$). MICHELLE
has a pixel scale of 0.1005'' per pixel. Imaging was taken using
a ABBA chop-nod sequence with a position angle of 30$^{\circ}$ for
all observations. The data taken under GN-2006A-Q-10 were divided into
3 observation groups, with observations of a standard star from \citet{1999AJ....117.1864C},
HD 98118 (M0III, F$_{18 \micron}$=4.2 Jy), taken before and after the
observations of $\beta$ Leo to serve as a standard star for flux calibration
and to monitor the PSF. The data taken under GN-2007A-C-10 were previously
published by \citet{2009ApJ...691.1896A}. Observations of  HD109511 (K0, F$_{18 \micron}=1.4$Jy) preceded and followed the observations of $\beta$ Leo as a standard star and PSF reference.
Combining all observations, the total on source time at 18 $\micron$
is 1751s and 1175s at 12 $\micron$. 

The data were reduced using custom routines described in \citet{2008AA...485..897S}.
The data reduction involved determining a gain map for each observation
using the mean values of each frame to construct a map of pixel responsivity.
The on-source pixels were masked during this process, making this
equivalent to a sky flat field frame. A DC offset was then determined
by calculating the mean pixel values in every row and every column,
again masking pixels where there was source emission present. This
was then subtracted from the final image to ensure a flat background.
Pixels which showed high or low gain in comparison with the median
response throughout the observation were masked off. To avoid
errors in co-adding the data which could arise from misalignment of
the images, we fitted a Gaussian with a sub-pixel centroid to accurately
determine the centre of the image and so the position of the star.
The re-binning was done using bilinear interpolation across the array.

The observations were divided into four groups (as shown in Table 1)
such that flux calibration for each group was done with standards
observed at a similar airmass to $\beta$ Leo. The calibration levels
were compared with the airmass for the standard star observations
and no correlation was found, so no extinction correction was applied
to the calibration factors.

For the images of $\beta$ Leo in each group a calibration factor was
determined using a co-add of the two standard star observations in
that group. The average calibration error for the 3 18 $\micron$ groups was
8 \%, but as seen in Table 1 the large calibration uncertainties lead to a wide range in fluxes among the three groups. The 12 $\micron$ group had a calibration error of 11\%.
These centred, flux calibrated images from each group were then used
to produce final images of $\beta$ Leo and the standard star. The first standard (Std1) and image (Im1) in Group 1 (See Table 1) at 18 $\micron$ and Std12.1 and Im12.1 at 12 $\micron$ showed elongation in
the telescope chop-nod direction, so were not used in the final co-added
images, but were used when calculating the flux calibration factors.

\begin{table*}
\caption{\label{tab:Observations-taken-under}Observations taken under proposal
GN-2006A-Q-10 and GN-2007A-C-10 in order. The integration time listed
is the on-source integration time. Fluxes are for a 1'' radius aperture
centred on the star. The group indicates the standard-science-standard
observing pattern used. }

\begin{tabular}{|c|c|c|c|c|c|c|c|}
\hline 
Program  & Date  & Object  & Group  & Name  & Filter  & Integration Time (s)  & Calibrated Flux (Jy)\tabularnewline
\hline 
GN-2006A-Q-10  & 10/05/06  & HD 98118  & 1  & Std1  & QA  & 82  & 4.20\tabularnewline
GN-2006A-Q-10  & 10/05/06  & $\beta$ Leo  & 1  & Im1  & QA  & 396  & $3.17\pm0.13$\tabularnewline
GN-2006A-Q-10  & 10/05/06  & $\beta$ Leo  & 1  & Im2  & QA  & 245  & $3.17\pm0.13$\tabularnewline
GN-2006A-Q-10  & 10/05/06  & HD 98118  & 1  & Std2  & QA  & 82  & 4.20\tabularnewline
\noalign{\vskip\doublerulesep} GN-2006A-Q-10  & 15/05/06  & HD98118  & 2  & Std3  & QA  & 82  & 4.20\tabularnewline
GN-2006A-Q-10  & 15/05/06  & $\beta$ Leo  & 2  & Im3  & QA  & 326  & $2.80\pm$0.25\tabularnewline
GN-2006A-Q-10  & 15/05/06  & HD 98118  & 2  & Std4  & QA  & 82  & 4.20\tabularnewline
\noalign{\vskip\doublerulesep} GN-2006A-Q-10  & 11/06/06  & HD 98118  & 3  & Std12.1  & N'  & 47  & 3.65 \tabularnewline
GN-2006A-Q-10  & 11/06/06  & $\beta$ Leo  & 3  & Im12.1  & N'  & 376  & $5.82\pm$0.51 \tabularnewline
GN-2006A-Q-10  & 11/06/06  & $\beta$ Leo  & 3  & Im12.2  & N'  & 376  & $5.82\pm$0.51 \tabularnewline
GN-2006A-Q-10  & 11/06/06  & $\beta$ Leo  & 3  & Im12.3  & N'  & 423  & $5.82\pm$0.51 \tabularnewline
GN-2006A-Q-10  & 11/06/06  & HD 98118  & 3  & Std12.2  & N'  & 47  & 3.65\tabularnewline
\noalign{\vskip\doublerulesep} GN-2007A-C-10  & 8/03/07  & HD109511  & 4  & Std5  & QA  & 86  & 1.4\tabularnewline
GN-2007A-C-10  & 8/03/07  & $\beta$ Leo  & 4  & Im4  & QA  & 392  & $2.36\pm$0.33\tabularnewline
GN-2007A-C-10  & 8/03/07  & $\beta$ Leo  & 4  & Im5  & QA  & 392  & $2.36\pm$0.33\tabularnewline
GN-2007A-C-10  & 8/03/07  & HD98118  & 4  & Std6  & QA  & 86  & 1.4\tabularnewline
\hline
\end{tabular}
\end{table*}

\begin{figure}
\caption{The profile of the line cuts through the total co-added Mid-IR images at
125$^{\circ}$, the position angle of the extension seen in the PACS
images. The dotted lines are the final co-added image of $\beta$ Leo
in each group and the solid lines are the final coadded standard star
in each group. The 12 $\micron$ line cut is the top image, the 18 $\micron$
is the bottom image. $\beta$ Leo shows no significant extension with respect
to the PSF, indicating the excess at this wavelength is unresolved. In fact, the observations appear narrower than the PSF, but this is not significant and is due to variation in observing conditions.\label{IR}}

\includegraphics[width=7cm, height=5cm]{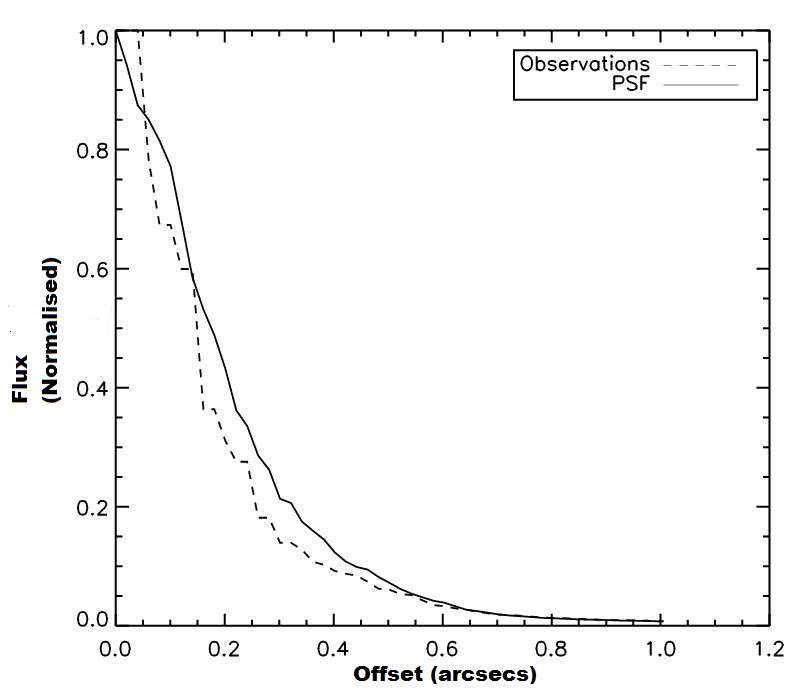}

\includegraphics[width=7cm, height=5cm]{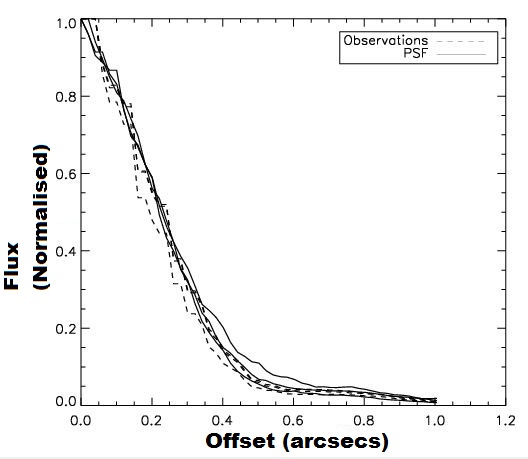}
\end{figure}

Photometry was performed on the final co-added images of $\beta$ Leo
using a 1.0 arcsec radius circular aperture centred on the star. The
stellar flux is expected to be 5297 mJy and 2020 mJy at 11.3 and 18.1 $\micron$ respectively,
from fitting a model spectrum as described in $\S$ 4. Our photometry yields
fluxes of 5822$\pm$476 mJy at at 12 $\micron$ and 2360$\pm$312 mJy
at 18.6  $\micron$ including both calibration and photometric errors,
giving an excess of 525$\pm$476 mJy at 12 $\micron$ and 340$\pm$312 mJy
at 18 $\micron$.  The statistical noise was determined using an annulus with
an inner radius of 7'' and an outer radius of 8'' centred on the
star, resulting in an error averaged over all the groups of 0.31 mJy/arcsec$^{2}$at
12 $\micron$ and 0.41 mJy/arcsec$^{2}$ at 18 $\micron$. The IRS spectrum presented in \citet{2010ApJ...724.1238S} gives
an excess above the photosphere of 61$\pm$103 mJy at 12 $\micron$ and 256$\pm$53 mJy
at 18 $\micron$. The photospheric fit used here is consistent with that used in \citet{2010ApJ...724.1238S}. The MICHELLE photometry presented here therefore agrees
with the IRS results, but the calibration errors are too large to detect the excess.
 To assess whether these mid-IR images have resolved the disc, line cuts were taken
 along  the PA of the extension (125\degr E of N)
seen in the PACS images (125$^{\circ}$)  for both $\beta$ Leo and the standard star . These are shown in Figure \ref{IR}. $\beta$ Leo shows no extension when compared with the PSF, and the excess
at both 12 and 18 $\micron$ is unresolved. This is consistent with the previous results of \citet{2009ApJ...691.1896A, 2010ApJ...530..329T}.  None of the groups show significant extension.
We also tried subtracting the standard star image scaled to the peak
flux of the $\beta$ Leo images, but no significant structure remains.

As the location of the disc is known from the resolved PACS images
with peak emission at $\sim$5'' radius, we tried convolving
the co-added images with a series of Gaussians with FWHMs from 1'' to 20'' 
to find any large scale, low surface brightness features in the outer
regions of the MICHELLE images. The MICHELLE field of view is 32''$\times$24'',
but the chop throw of 15'' limits the usable region. We find no
coherent features at the expected location of the debris disc in the
convolved image. The surface brightness
in an annulus from 0.5'' to 1'' in the MICHELLE 18 $\micron$ imaging of $\beta$ Leo is 0.41$\pm$0.34 mJy/pixel.
However, the brightness in the same annulus of the PSF scaled to the
flux of $\beta$ Leo is 0.31 mJy/pixel. 

\subsection{SCUBA 2}

\begin{figure*}
\caption{Spitzer MIPS observations of $\beta$ Leo at 24 $\micron$. The disc is detected and partially resolved compared to the PSF. The disc image is shown in the left panel, the PSF is the centre panel and the lower panel shows the surface brightness profiles of the Observations (solid line) and the PSF (dashed line) showing the filling in of the first dark Airy ring between 5.5\arcsec and 7.5 \arcsec.  \label{MIPS} }
\subfigure[Spitzer MIPS 24 $\micron$ Observations of $\beta$ Leo.]{\includegraphics[width=6.5cm]{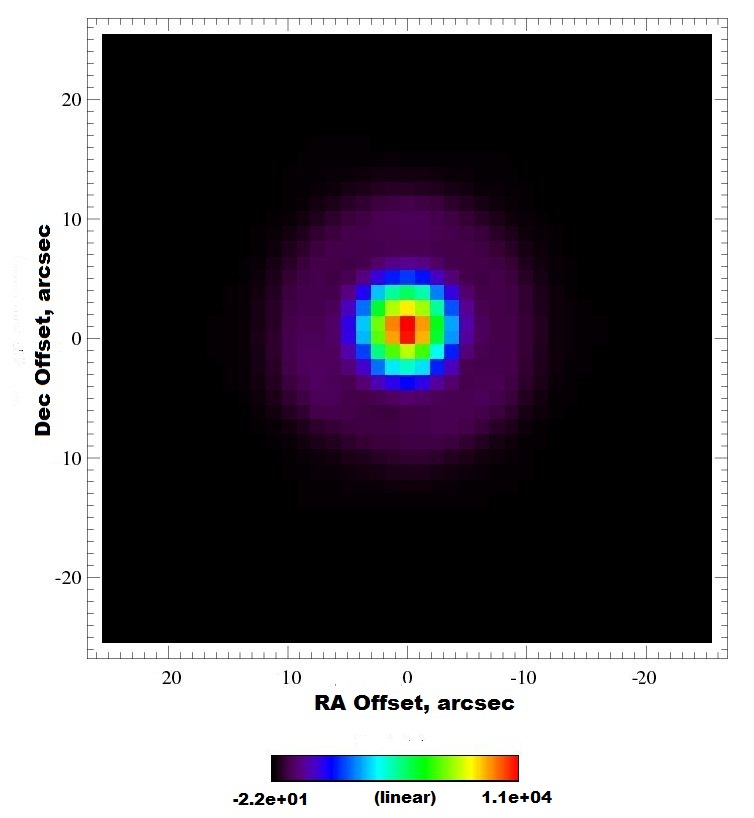}}\subfigure[Spitzer MIPS 24 $\micron$ PSF]{\includegraphics[ width=6.5cm]{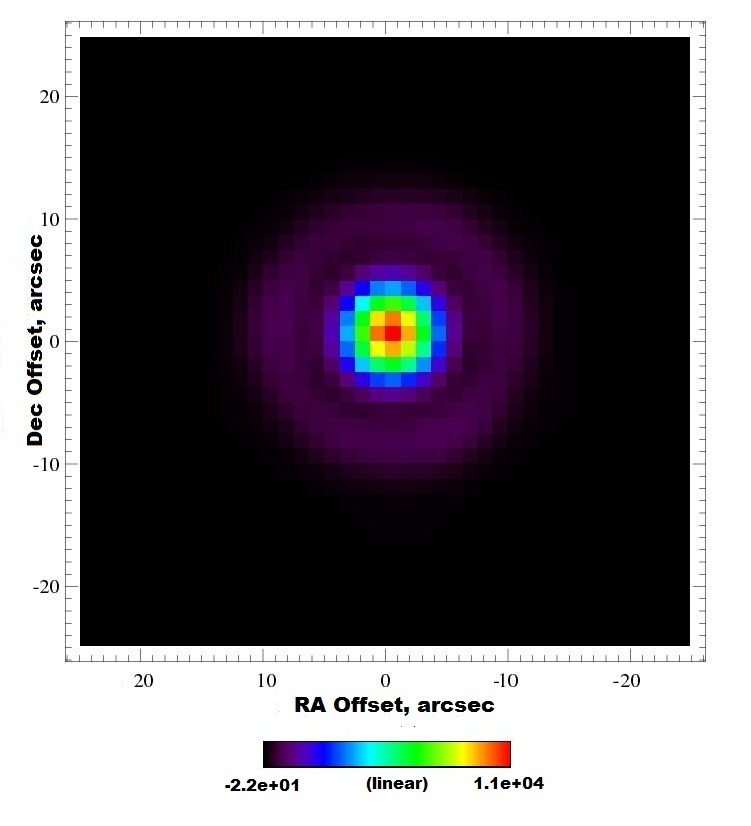}}\vspace{0.3cm}
\centering\subfigure[Surface brightness profile of 24$\micron$ $\beta$ Leo observation and the PSF]{\includegraphics[ width=6.5cm, height=6.5cm]{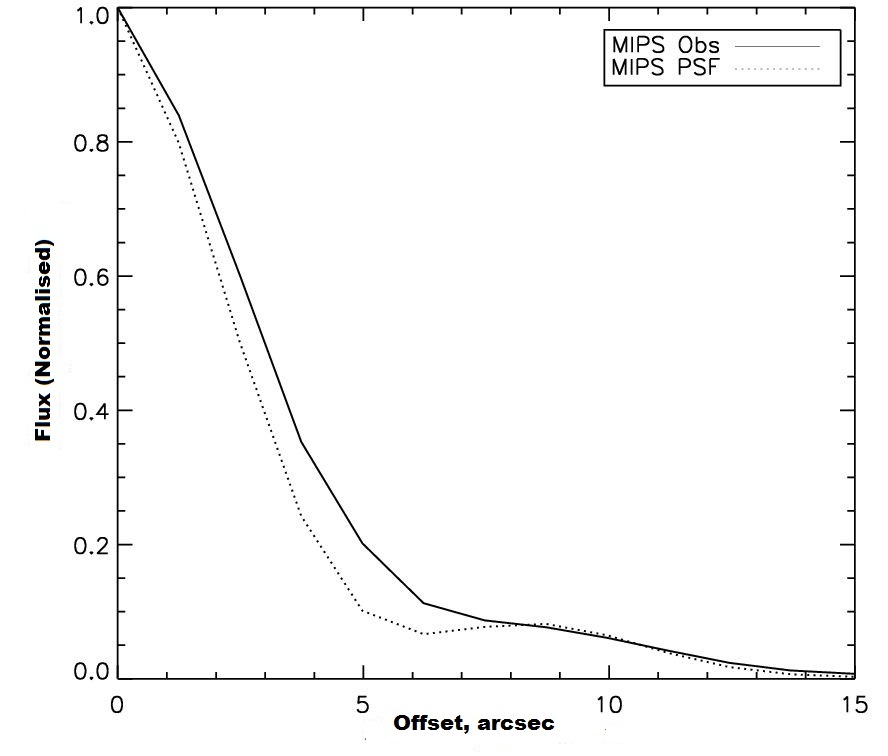}}
\end{figure*}

The SCUBA 2 instrument  (Sub-mm Common User Bolometer Array 2) (\citealp{2006SPIE.6275E..45H}) operating on the JCMT ((James Clerk Maxwell Telescope) was used to obtain fully sampled maps of $\beta$ Leo, and the surrounding region at 450\,$\mu$m and 850\,$\mu$m.  These data were obtained on the nights of the 16th and 17th February 2010 as part of the shared risk observing phase of instrument commissioning.  The final maps contain data from six observations obtained on each night, giving a median integration time of 3 hours per pixel in the central one square arcminute, and a total observing time of 6 hours.  The `daisy' pattern observing mode was used at a scanning rate of 120 arcseconds per second (\citealp{2010SPIE.7740E..66K}), giving a map with a usable area of $\sim$9 square arcminutes.

The data were processed and calibrated using the SMURF package in the Namaka Starlink release (\citet{2010arXiv1011.5876J,2010SPIE.7741E..54D}).  The data were high-pass filtered to mitigate the 1/f noise present in the data, with filter parameters set so as to retain features smaller than $\sim$120 arcseconds.
Maps were obtained at 450 $\micron$ and 850 $\micron$ and the data were reduced using the latest version of the SMURF pipeline. $\beta$ Leo was not detected in either band.  Photometry on the 450 $\micron$ map gives a 3 $\sigma$ upper limit of 50 mJy/beam At 850 $\micron$ the 3 $\sigma$ upper limit is 6 mJy/beam. The expected stellar flux in these wavebands is 3.1 mJy and 0.8 mJy respectively. \citet{2003AJ....125.3334H} found an upper limit of 20 mJy at 870$\micron$, so these upper limits significantly improve the constraints on the SED in the sub-mm and were used in the SED fitting described in \S\ref{Model}.

\subsection{Spitzer MIPS Observations}

In addition to the photometric points at 24 and 70 $\micron$ on the SED shown in Figure 1, we also include the partially resolved MIPS image of the disc at 24 $\micron$, presented in \citet{2010ApJ...724.1238S}. \citet{2010ApJ...724.1238S} observed that the 24 $\micron$ photometric point obtained using an aperture of 14\arcsec.94 gives a total integrated flux of 1623$\pm$33 mJy which is 2.5\% higher than the flux obtained in a 6.\arcsec23 aperture, leading to the suggestion that the disc may be marginally resolved at this wavelength. The photometric value from the larger aperture is used for SED fitting. The MIPS 24 $\micron$ image is shown in the first panel of Figure \ref{MIPS}. Compared to the PSF shown in the right panel of Figure \ref{MIPS} the first dark Airy ring (between radii of 5.5\arcsec to 7.5\arcsec) appears more filled in. Subtraction of the PSF scaled to the expected photospheric flux of $\beta$ Leo gives residuals with a FWHM of 6.88\arcsec $\times$6.61\arcsec at a position angle of 118$^{\circ}$ which is larger than the FWHM of the PSF (5.69\arcsec$\times$5.53"). This is again consistent with a disc that is marginally extended with an inclination of 60$\pm$10$^{\circ}$ from edge on and a position angle of 118$\pm$3$^{\circ}$. Comparing $\beta$ Leo to the PSF observed for a star with hot dust at <2 AU (i.e. similar to the inteferometric detection) of the same spectral type ($\zeta$ Lep - 5.$\arcsec$61$\times$5.$\arcsec$55 \citealp{2008ApJ...679L.125S}) and to a theoretical PSF (generated using TinyTim \footnote{Tiny Tim/Spitzer, developed by John Krist for the Spitzer Science Center. The Center is managed by the California Institute of Technology under a contract with NASA})  also suggests that the disc is marginally resolved.  If the 24 $\micron$ observations are deconvolved from the PSF (i.e. the FWHM of the PSF is subtracted in quadrature from the FWHM of the observations), this gives the FWHM of the residuals as 3.9$\arcsec$. The position angle and inclination of these residuals are consistent with the disc resolved using PACS. The 70 $\micron$ Spitzer MIPS obsevations are consistent with a point source. We include the 24 $\micron$ image when considering the observational constraints for a model of the $\beta$ Leo debris disc.

\subsection{Optical Observations}
\label{ACS_obs}
$\beta$ Leo was imaged using the Hubble Space Telescope Advanced Camera for Surveys High Resolution Channel  (HRC) coronagraph on March 25, 2004 (GO-9475, PI Kalas).  We occulted $\beta$ Leo with the 1.8$"$ diameter occulting spot approximately centered on the 1024$\times$1024 pixel camera.  Seven 140 second exposures were acquired using the F606W filter ($\lambda_{c} = 591$ nm, $\delta\lambda = 234$ nm).  The A1V star HD95418 was observed before $\beta$ Leo to serve as a reference for the stellar PSF .  Cosmic rays were filtered by taking the median combination of the seven $\beta$ Leo exposures, as well as the six 180 second exposures on HD95418. It should be noted that HD95418 $\beta$ UMa has an IR-excess indicative of a debris disc but this is not detected in scattered light. The HD 95418 observation was registered and scaled by a factor of 1.22 to subtract the $\beta$ Leo PSF. The resulting, PSF-subtracted image was then corrected for geometric distortion giving a pixel scale of 0.025$\arcsec$ pix$^{-1}$.  Figure 5 shows the resulting optical image of $\beta$ Leo.  We find a residual halo of light in an annulus $2.5-4.0\arcsec$ from the star.  This halo could plausibly originate from either dust scattered light or an imperfect PSF subtraction.  The latter effect results from thermal ``breathing'' of the telescope between the observations of the two stars, as well as a small color mismatch between $\beta$ Leo and HD 95418. The limits on surface brightness from these observations were used as constraints for the Modelling described in \S\ref{Model}.

\subsection{Herschel SPIRE Observations}

 Photometric observations of $\beta$ Leo were obtained with the SPIRE instrument on Herschel (\citealp{2010AA...518L...3G}), providing maps in wavebands centred at 250, 350, and 500 $\mu$m.  These data were observed on the 23rd November 2011 (Herschel operational day 558) using the `small map' observing mode.  Five repeat maps were performed, effectively giving  confusion noise limited maps in all bands.  The data were reduced using HIPE Version 6.0, build number 1985, and the standard pipeline script was used to perform the reduction. The images are shown in Figure \ref{spire}.  Photometry was performed on the images. At 250 $\micron$ a 14\arcsec radius circular aperture  (beam size 18.9\arcsec$\times$17.6\arcsec \footnote{SPIRE Operating Manual: \url{http://herschel.esac.esa.int/hcss-doc-5.0/print/spire_um/spire_um.pdf}})  gives 51$\pm$12 mJy. At 350 $\micron$ a 30\arcsec aperture (beam size 25.6\arcsec$\times$24.2") gives 18$\pm$7 mJy which gives a 3$\sigma$ upper limit of 39 mJy. At 500 $\micron$ a 40\arcsec aperture (beam size 38.0\arcsec$\times$34.6\arcsec) gives 2.8$\pm$4.1 mJy, which corresponds to a 3$\sigma$ upper limit of 15.1 mJy.

To assess whether the 250 $\micron$ image had resolved the disc we  fit the image with a 2D Gaussian, which has a FWHM of 22.5\arcsec$\pm$1.2$\times$18.7\arcsec$\pm$1.1 with a PA of 118$^{\circ}$, consistent with the PA of  the resolved disc (125$\pm$15$^{\circ}$) from the PACS images. Comparison of the disc size with that of the SPIRE PSF (Neptune), found no significant extension at 250 $\micron$.

\begin{figure}
\caption{Optical HST observations of Beta Leo after PSF subtraction.  North is up, east is left.  The green circle has diameter 8$\arcsec$ or 88 AU.  The residual halo of light discussed in the text is contained interior to the green circle.  Saturation columns are evident to the upper right and lower left of the star.  The bright circular structure above the saturation columns is a different occulting spot in the HRC focal plane. \label{ACS}}
\includegraphics[width=8cm]{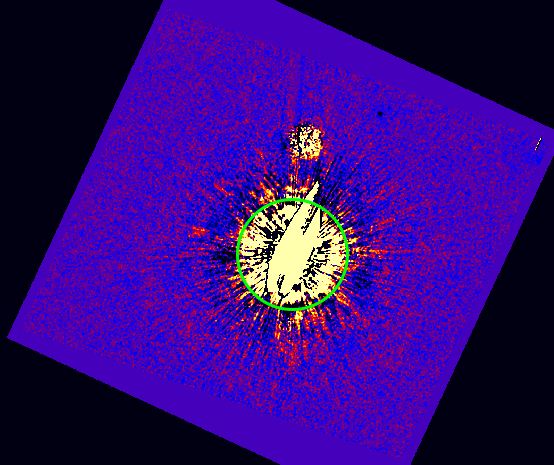}
\end{figure}
\begin{figure}

\caption{SPIRE observations of $\beta$ Leo at 250 $\micron$. The disc is detected but not resolved at 250$\micron$. At 350 there is a 2$\sigma$ detection, giving a 3 $\sigma$ upper limit on the disc and star emission of 21 mJy and the non-detection at 500 $\micron$ also gives an 3 $\sigma$ upper limit of 12.3 mJy, which is  consistent with the fluxes expected from the SED modelling.\label{spire} }

\centering \includegraphics[bb=100 60 625 625, width=7cm]{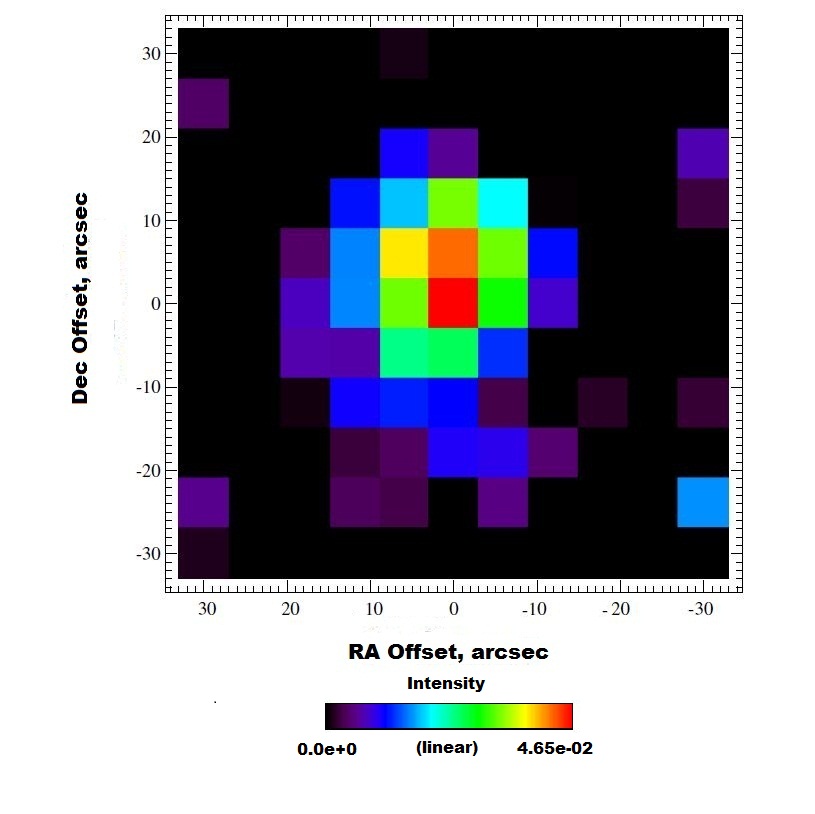}
\end{figure}

\subsection{Interferometry \label{hot} }

\citet{2009ApJ...691.1896A} present interferometric observations of
$\beta$ Leo at 2 $\micron$ using the FLUOR instrument at the CHARA interferometer. These data show a short-baseline visibility deficit, indicating that the
source is somewhat extended. Akeson et al. (2009)  suggest a model in which the inferred few percent
NIR (Near Infra-red) excess is due to a population of dust grains within the field
of view (<4.6 AU), comprised of a population of blackbody grains that extends
inward to the sublimation radius (0.12 AU for $\beta$ Leo for a sublimation
temperature of 1600K), which maximises the 2 $\micron$ emission. This
places a lower limit on the fractional dust luminosity (defined as L$_{IR}$/L$_{*}$ i.e. a measure of the IR excess for each component.)
for this hot population of 8.0$\pm$1.1$\times$10$^{-5}$, which
is larger than the value of 2.7$\times$10$^{-5}$ for the IRS
emission (\citealp{2006ApJS..166..351C}). The strongest spatial constraint
from the interferometry is that there is dust with a fractional luminosity of 8$\times$10$^{-5}$ within 4.6 AU.

 Stock et al. (2010) present new 10 $\micron$ nulling interferometry data from the Keck Interferometer Nuller (KIN), which suggest that there is no significant resolved emission
at 10 $\micron$ in the KIN beam (FWHM of 0.50\arcsec $\times$ 0.44\arcsec \citep{2009PASP..121.1120C}). Limited by its small field of view (0.6\arcsec), these observations are
unable to detect extended structures beyond a radius of 3 AU at the distance of $\beta$ Leo
(11 pc). Moreover, due to its complex transmission function, KIN observations are most
sensitive to extended structures with radii from 0.1-1 AU. Stock et al. (2010) also present  an N-band (8-13 $\micron$ ) nulling interferometry detection with BLINC of a null of 1.74$\pm$0.3\%, which leads to a model-based estimate of an excess flux of 250+/-50 mJy. BLINC is sensitive to the region between 1 and 9 AU (between 0.12 and 0.8$\arcsec$). Stock  et al. (2010) quote limits for various uniform disc models: 110 mJy for a disc between 0 and 1 AU, 240 mJy for a disc from 0-2 AU and 370 mJy for  a 1 to 2 AU ring. Stock et al (2010)'s preferred model is a ring from 2-3 AU with a flat surface density profile, and a flux of 250$\pm$50 mJy at 10 $\micron$. This flux is used as a constraint in the SED modelling. 

\section{Modelling}\label{Model}

\begin{table*}
\caption{Table listing the data used in the modelling process, indicating if it was used to constrain the SED or the images or both. Where the table indicates that the observations were used as a surface brightness constraint this indicates that model images were produced at this wavelength to check that the predicted model surface brightness was compatible with the observed limits. \label{data}}
\begin{tabular}{|c|c|c|c|c|c|c|}
\hline 
Wavelength/Waveband & Flux (Jy) & Error & Instrument/Survey & Resolved & Constraint on Modelling & Reference\tabularnewline
\hline
\hline 
1.6 $\micron$ & N/A & N/A & HST ACS & No & Image Surface Brightness (Upper Limit) & $\S$ 2.6\tabularnewline
3.6 $\micron$ & 49.435 & 1.085 & Spitzer IRAC & No & SED & \citet{2010ApJ...724.1238S}\tabularnewline
4.5 $\micron$ & 30.971 & 0.693 & Spitzer IRAC & No & SED & \citet{2010ApJ...724.1238S}\tabularnewline
5.8 $\micron$  & 20.450 & 0.458 & Spitzer IRAC & No & SED &\citet{2010ApJ...724.1238S}\tabularnewline
6.75 $\micron$ & 15.07 & 0.081 & Spitzer IRS & No & SED & \citet{2006ApJS..166..351C}\tabularnewline
8.0 $\micron$ & 10.759 & 0.244 & Spitzer IRAC & No & SED & \citet{2010ApJ...724.1238S}\tabularnewline
9.0 & 8.094 & 0.096 & AKARI & No & SED & \citet{2007PASJ...59S.369M}\tabularnewline
9.5 $\micron$ & 8.22 & 0.082 & Spitzer IRS & No & SED & \citet{2006ApJS..166..351C}\tabularnewline
10.5 $\micron$ & 6.55 & 0.05 & BLINC & Yes & SED & \citet{2010ApJ...724.1238S}\tabularnewline
11.3$\micron$ (N band) & 5.82 & 0.48 & Gemini MICHELLE & No & Surface Brightness limits, SED & $\S$2.3\tabularnewline
 13. 0 $\micron$ & 4.29 & 0.024 & Spitzer IRS & No & SED & \citet{2006ApJS..166..351C}\tabularnewline
15.5 $\micron$ & 3.36 & 0.073 & Spitzer IRS & No & SED & \citet{2006ApJS..166..351C}\tabularnewline
18.1 $\micron$ (Q Band) & 2.36 & 0.31 & Gemini MICHELLE & No & Surface Brightness limits, SED & $\S$2.3\tabularnewline
20.0 $\micron$ & 2.30 & 0.021 & Spitzer IRS & No & SED & \citet{2006ApJS..166..351C}\tabularnewline
24 $\micron$ & 1.647 & 0.033 & Spitzer MIPS & Marginally & SED, Image & \citet{2010ApJ...724.1238S}\tabularnewline
 29. 0 $\micron$ & 1.68 & 0.009 & Spitzer IRS & No & SED & \citet{2006ApJS..166..351C}\tabularnewline
33.5 $\micron$ & 1.54 & 0.014 & Spitzer IRS & No & SED & \citet{2006ApJS..166..351C}\tabularnewline
 70 $\micron$ & 0.743 & 0.052 & Spitzer MIPS & No & SED & \citet{2010ApJ...724.1238S}\tabularnewline
100 $\micron$ & 0.480 & 0.030 & Herschel PACS & Yes & SED, Image & $\S$2.2\tabularnewline
160 $\micron$ & 0.215 & 0.032 & Herschel PACS & Yes & SED, Image & $\S$2.2\tabularnewline
250 $\micron$ & 0.051 & 0.012 & Herschel SPIRE & No & SED, Image & $\S$2.7\tabularnewline
350 $\micron$ & <0.039 & N/A & Herschel SPIRE & No & SED upper limit & $\S$2.7\tabularnewline
450 $\micron$ & <0.050 & N/A & SCUBA-2 & No & SED upper limit & $\S$2.4\tabularnewline
500 $\micron$ & <0.015 & N/A & Herschel SPIRE & No & SED upper limit & $\S$2.7\tabularnewline
850 $\micron$ & <0.006 & N/A & SCUBA-2 & No & SED upper limit & $\S$2.4\tabularnewline
870 $\micron$ & <0.20 & N/A & HHT Sub-mm Observatory & No & SED upper limit & \citet{2003AJ....125.3334H}\tabularnewline
\hline
\end{tabular}
\end{table*}

To determine constraints on the radial distribution of the emission
seen in the resolved 100 and 160 $\micron$ images we first considered a
model of the disc structure. For example in $\S$3.1 we consider the simplest possible model. This model is composed of a single
axisymmetric disc component, defined by four free parameters: inner
radius (r$_{in}$), outer radius (r$_{out}$), inclination (i) and surface density distribution
$\Sigma$, which is assumed to have the form $\Sigma\propto r^{\gamma}$.
The disc opening angle (which sets the disc height) is assumed to
be 5$^{\circ}$ but this parameter is unconstrained by the modelling
process. The flux from annuli in the disc at different radii was determined
assuming a grain composition and size distribution that were constrained
using the emission spectrum.

In order to constrain the dust location, the width of the
disc, the inclination and the surface density profile, a grid of models
was run. The data used in this modelling is listed in Table \ref{data}, which indicates which observations were used to constrain the SED and the images. The model images were convolved with a PSF (image of $\alpha$ Boo taken in the same observing mode as the observations
of $\beta$ Leo and reduced in the same way) and compared with the observations using the images as well as the linecuts both in the direction
of extension (PA $125^{\circ}$) and perpendicular to the extension
(PA $35^{\circ}$). The use of both linecuts allowed us to constrain
the inclination simultaneously with the radial morphology. Both wavelengths
were fit simultaneously, and a joint best fit was determined by
minimising the combined reduced $\chi^{2}$. The reduced $\chi^{2}$
of the fit ($\chi_{r}^{2}=\frac{(obs-mod)^{2}}{\nu}$ where $\nu$
is the number of free parameters and 1 represents a good fit) to each
of these six pieces of observational data were calculated and were
then combined linearly with equal weight to come to a final 
chi-squared ($\chi_{r_{res}}^{2}=\chi_{rline_{125}}^{2}+\chi_{rline_{35}}^{2}$). 

The emission spectrum of the photosphere of  $\beta$ Leo was discussed in $\S$ 2.1. To calculate the emission from the dust grains in the model they were assumed
to have a size distribution with $n(D)\propto D^{-3.5}$, where D
is the grain diameter, which is the standard solution for a theoretical
collisional cascade (see \citealp{1969JGR....74.2531D}), that is truncated
at a minimum and maximum grain size. The minimum grain size was treated as a free parameter with a range between 0.01-100 times $D_{bl}$, where $D_{bl}$ is the largest grain size blown out of the system by radiation pressure and depends on grain composition and stellar properties. The maximum size was fixed at 1cm since larger grains have a negligible
contribution to flux from a model with this size distribution. The
grains were assumed to have a silicate core (amorphous olivine) and
an accreted mantle of organic refractories produced by UV photoprocessing
of ice (as used in \citet{1997AA...323..566L,1999AA...348..557A}).
A range of compositions was also tried, with amorphous silicate fractions
varying from 0\% to 90\% by volume and with porosities
(i.e. vacuum fraction of grain by volume) from 0 \% to 95 \%.  Dielectric constants were calculated from tabulated laboratory
values (\citealp{1997AA...323..566L,1999AA...348..557A}) using Maxwell-Garnett
effective medium theory. The optical properties of the grains were
calculated using Mie theory, Rayleigh-Gans theory and Geometric Optics
in the appropriate size regimes (\citealp{1983asls.book.....B}). This composition was the best fit found by minimising $\chi^{2}$ across a grid representing the possible compositions. However, we do not give constraints on the composition because the model used contains uncertainties in the calculation of the optical properties and we do not want to overempahise this aspect of the modelling.

These models were fit to the IRS spectrum,  BLINC point,  MIPS points, PACS points and SCUBA-2 upper limits. Table  \ref{data}  lists the data used in the modelling and indicates if it was used to constrain the SED, images or both. As the IRS spectrum has no obvious features to fit to,
to calculate $\chi^{2}_{SED}$ we chose 6 windows in the spectrum, each
$\sim$1-2 $\micron$ wide, with a constant signal to noise ratio
in the window. This gives 11 fluxes (6 IRS, 1 BLINC, 2 MIPS, 2 PACS) and 2
upper limits with which to calculate the $\chi^{2}_{SED}$. 

The IRS spectrum is $\sim$2\% lower than the MIPS observation at 24 $\micron$ as the MIPS observation is marginally resolved, as discussed in $\S$ 2.5. When calculating the MIPS 24 $\micron$ photometry a larger aperture was used to ensure all the flux was included, whereas the IRS spectrum assumes a point source.  A synthetic model image at 24 $\micron$ was compared with the MIPS image to check that the model reproduces the marginal extension observed. The total flux was compared to both the IRS and MIPS points and the $\sim$2\% discrepancy is a small fraction of the total flux and does not affect any conclusion about the goodness of fit of the models. There is a similar discrepancy between the BLINC flux at 10 $\micron$ and the IRS spectrum, which is again small enough not to affect the conclusions of the modelling within the uncertainties.


We considered models of increasing complexity. The initial model (described in $\S$ 3.1) is a 2 component model with a hot inner disc and a cold outer disc that has been resolved with PACS. The second model has 3 components, a hot inner disc, a warm component and a cold outer disc and is described in $\S$ 3.2. The third model consists of a single eccentric planetesimal population and is described in $\S$ 3.3.

\subsection{ 2 Component Model \label{best fit}}

\begin{figure*}
\caption{Spectral Energy Distribution (SED) of $\beta$ Leo. The photosphere of
$\beta$ Leo is fitted with a Kurucz model profile (L$_{star}$=14.0L$_{\odot}$, T$_{star}$=8660K) fitted to the 2MASS
fluxes and shown with a dark solid line. The IRS spectrum  (solid line from 5-30 $\micron$) of
\citealp{2006ApJS..166..351C}, MIPS
fluxes  (crosses), BLINC 10.1$\micron$ (upright cross) and Gemini MICHELLE fluxes in 1'' radius aperture
form imaging described in\S2.2 (diamonds) and PACS fluxes in 20''
aperture from imaging described in \S1 (squares). Upper limits are 3- $\sigma$ from SPIRE (350, 500 \micron), SCUBA-2 (450 and 850 \micron), SCUBA (higher 450 $\micron$, 850 $\micron$  values) and Bolocam (1.1mm), The excess is fitted
with a 2 component  realistic grain fit described in \S3.2 (dark solid line - hot component is shown with a dashed line and cold component is shown with a dotted line).\vspace{0.5cm}\label{SED_real}}

\includegraphics[width=11.0cm]{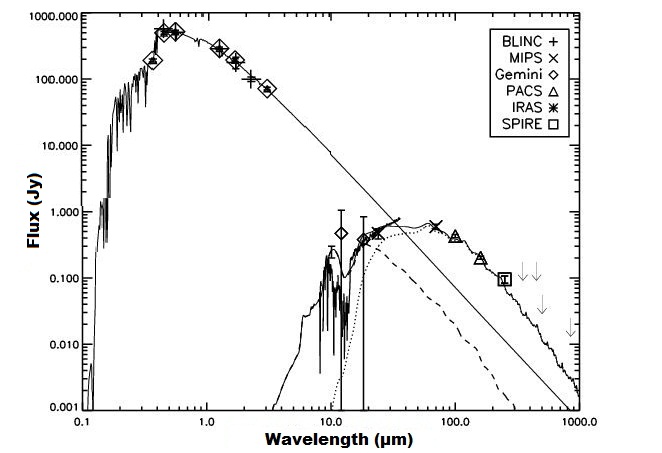}
\vspace{0.75cm}
\end{figure*}

\begin{figure*}

\caption{Model PACS images for best fit 2 component model - a cold disc from 15-70 AU with a surface density profile $\Sigma \propto r^{-1.5}$ and an unresolved hot component at 2 AU. This is shown to the same colour scale and pixel size (1\arcsec at 100 and 2\arcsec at 160 $\micron$) as the observations shown in Figure \ref{PACS}. The 100 $\micron$ image is on the left, the 160 $\micron$ is on the right. The residuals (observations -model) are below, with 100 on the left and 160 $\micron$ on the right respectively.\label{PACS_mod}}

\includegraphics[width=6cm]{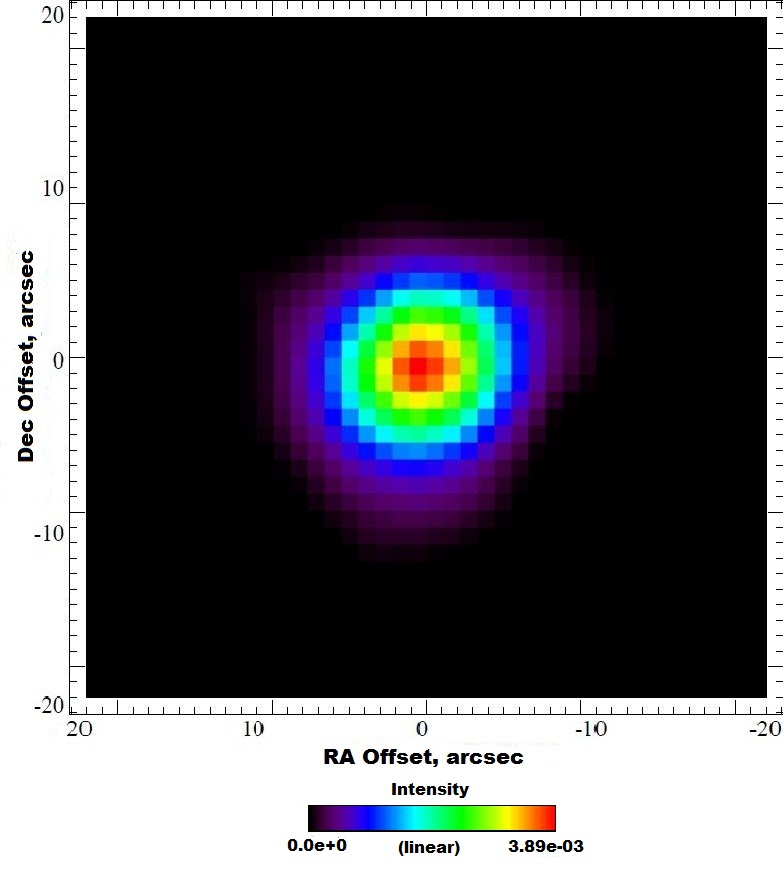}\hspace{0.4cm}\includegraphics[width=6.2cm]{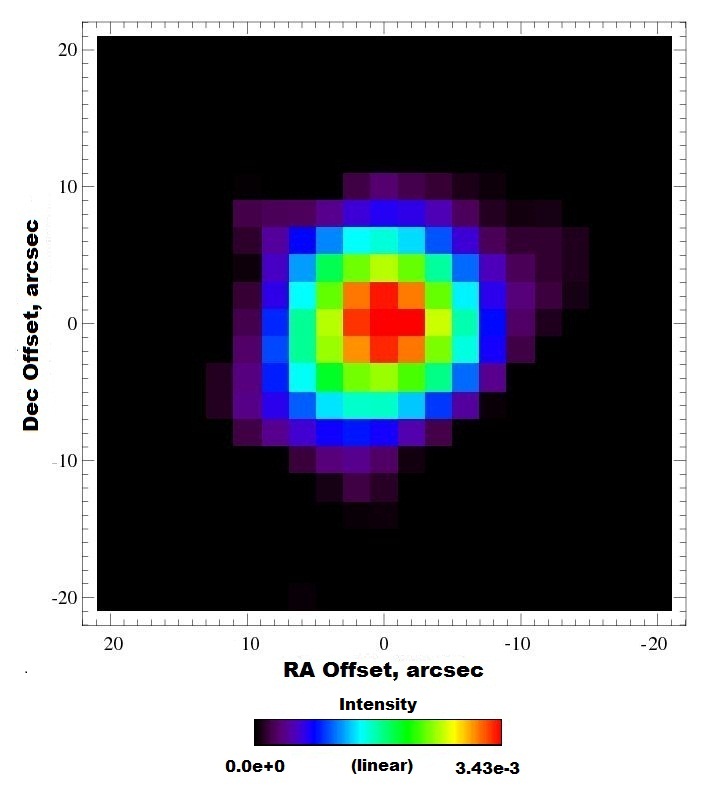}
\includegraphics[width=6cm]{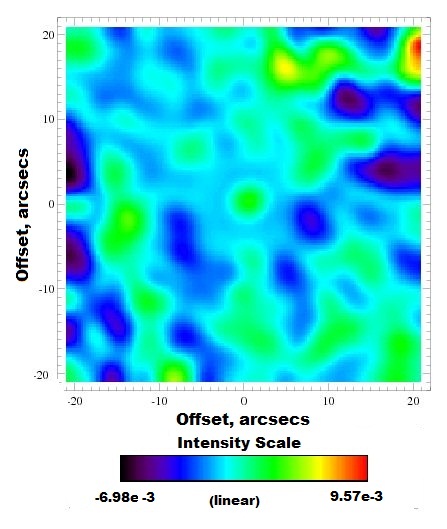}\hspace{0.4cm}\includegraphics[width=5.8cm]{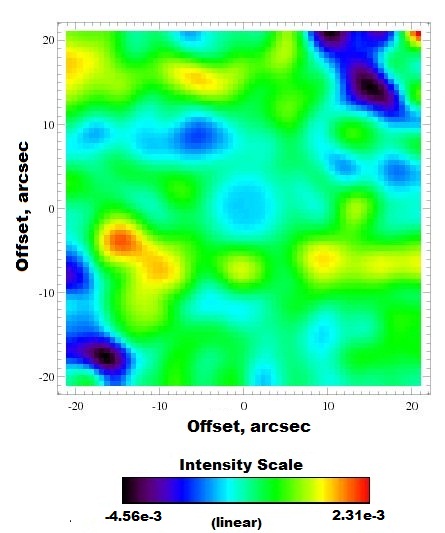}
\end{figure*}

 \label{composition}

The ranges of model parameters tested for the 2 component model were $R_{in}$:5 to 80 AU (1 AU intervals), $R_{out}$:
20 to 150 AU (5 AU intervals), inclination: 0$^{\circ}$ to 90$^{\circ}$
($5{}^{\circ}$) intervals, where 0$^{\circ}$ is edge on), surface
density index $\gamma$: 0 to -3.0 (0.5 intervals). The values
for gamma were chosen to cover possibilities such as the surface density
distribution expected from grains being blown out of the system by
radiation pressure ($\gamma$=-1.0), and that of the Minimum Mass
Solar Nebula (MMSN) ($\gamma$=-1.5).
The best fit 2 component model to the PACS 100 and 160 $\micron$ images, when considered iteratively with the SED fitting described in \S 3 was found to be a ring between 15$\pm10$ to 70$\pm5$ AU, with an inclination
of 55$\pm$5$^{\circ}$ from edge on, and a surface density profile index
$\gamma$=-1.5$\pm0.5$.  minimum grain diameter of 0.5$\times D_{bl}$
(3 $\micron$), a fixed  maximum grain size of 1cm and a composition with
a silicate fraction of 20 \%, a porosity of 20 \% with the rest of the
grain composed of organic refractories, with no ices present. The hot dust is assumed to lie from 2-3 AU with a flat surface density profile with a silicate fraction of 60\% and a porosity of 20\% and a minimum grain diameter of 0.6 $\micron$.  
 
The fractional luminosity of the cold component is 3$\times$10$^{-5}$. As there are no obvious features in the IRS spectrum, the main constraint on the composition
and size distribution comes from allowing the appropriate range of
temperatures to be present, given the constrains that the grains are
in the region 15-70 AU; blackbody grains would have to be at a radius
of 19 AU to achieve the observed temperature of 120 K. Although our model provides
a consistent fit to both image and SED, it is not expected that the
composition has been uniquely constrained by this process.

 To evaluate the fit of the model to both the SED and the images, we use $\chi^{2}_{r_{combined}}$=$\chi^{2}_{rres}+\chi^{2}_{SED}$, where $\chi^{2}_{rres}=\chi_{rline_{100}}^{2}+\chi_{rline_{160}}^{2}$ and $\chi^{2}_{r_{combined}}$ is the combined reduced $\chi^{2}$. For a model that was a perfect fit to the observations the reduced $\chi^{2}$ would be 1. This best fit model had $\chi_{r_{combined}}^{2}$=4.5
with $\chi_{rline_{100}}^{2}$=1.46, $\chi_{rline_{160}}^{2}$=1.58 and $\chi_{SED}^{2}$=1.46. For a perfect fit, $\chi^{2}_{r_{combined}}$ would be 3.
The best fit $\chi_{r_{combined}}^{2}$ model image is shown on the left of
Figure \ref{PACS_mod} and can be compared directly with the observed disc (Figure \ref{PACS}) which has the same colour scale.

\begin{figure} 
\caption{The top figure shows Bayesian Marginalisation of the best-fit model grid showing the probability distribution for the inner radius fit The bottom figure shows Bayesian Marginalisation of the best-fit model grid showing the probability distribution between the outer radius and the power law index of the surface density distribution, gamma. The highest probability is represented by the lightest colour, with a maximum probability of 0.53. \label{r_in_bayes}.}
\includegraphics[width=7.3cm]{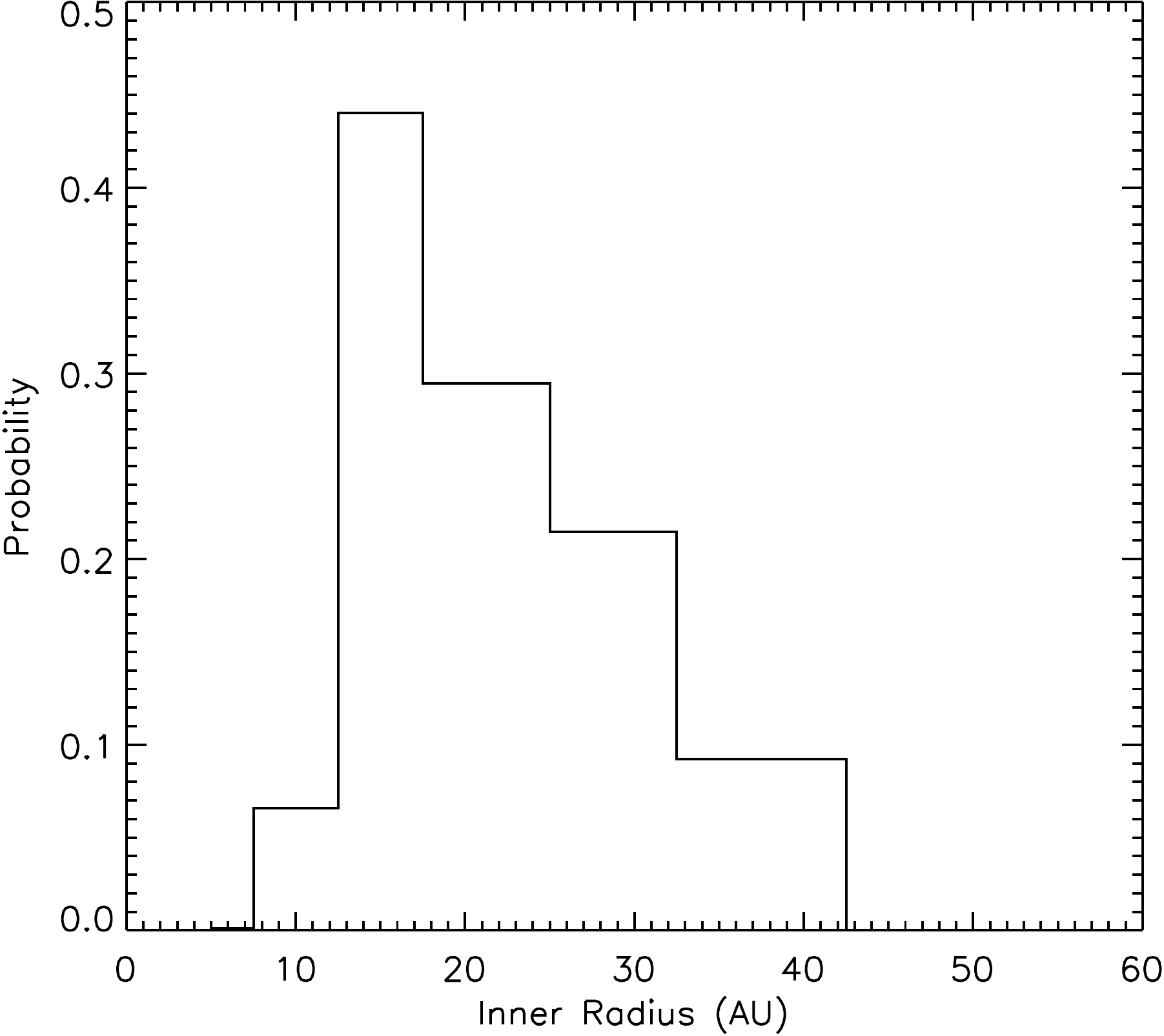}
\includegraphics[width=7.3cm]{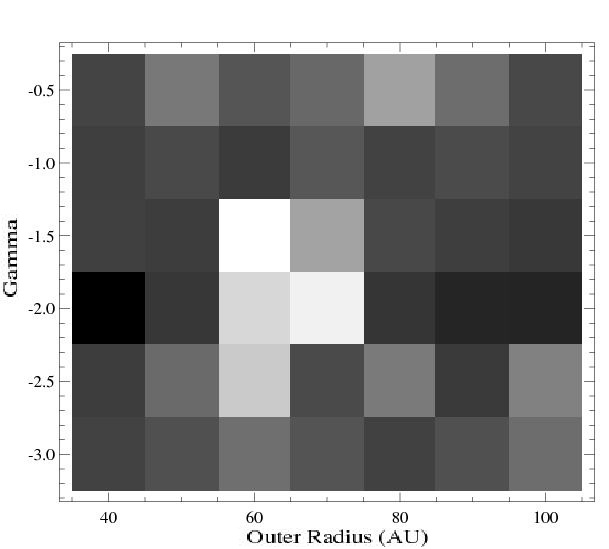}
\end{figure}

There was a difference between the models preferred by the SED and the images. The SED best fit requires a closer inner edge (r$_{in}$=10 AU) to produce enough flux in the 10-70 $\micron$ region whereas the images prefer a slightly larger inner radius (r$_{in}$=25 AU) with the same size distribution and grain composition.The best fit value of r$_{in}$ is therefore a compromise between these two values. To constrain the error on the inner radius, a Bayesian inference method is adopted (e.g., \citealp{1997ApJ...489..917L}, \citealp{2002ApJ...566.1124A}, \citealp{2008AA...489..633P}), in which each model is assigned a probability that the data are drawn from the model parameters. In cases where the Bayesian prior has a uniform probability distribution, as is the case here, this probability  is P=$P_{0}exp^{-\chi^{2}/2}$, where $\chi^{2}$ is the unreduced chi squared. The normalisation constant, P$_{0}$, is chosen so that the sum of the probabilities over all models in the grid is unity. Once this is done for all models in the grid, the  probability distribution for a given parameter can be derived by marginalising the 8-dimensional probability hypercube against the other 7 dimensions. An example of this process for the disc inner radius is shown in Figure \ref{r_in_bayes}, which shows that the distribution for possible locations of the inner edge peaks at 15 AU with a range of 10-40 AU. 

It is also possible to marginalise 2 parameters against the other six, creating a 3D plot with the probability distribution which allows us to examine the dependence of parameters upon each other. Figure \ref{r_in_bayes} also  shows such a plot of the marginalisation for the outer radius and gamma, the exponent of the surface density power law. The outer radius and surface distribution  are not well constrained by fitting the SED alone, partly due to the lack of constraints at wavelengths longer than 160 $\micron$. The images suggest that the outer radius required is >60 AU, but these parameters are degenerate as the steeper the surface density profile, the larger the outer radius needed to provide a good fit. The best fit choice of r$_{out}$=70 AU and surface density index $\gamma$=-1.5 represent the best compromise between these 2 factors. The grain size constraints on composition and minimum grain size come solely from the SED.

Compared to the two-component disc model presented in Stock et al. (2010), which has a ring of 1 $\micron$ diameter carbonaceous grains in a ring at 2-3 AU and a disc from 5-55 AU with a flat surface density profile ($\gamma$=0)  composed of silicate grains with radii from 5-1000 $\micron$. The Stock et al. (2010) model is driven by the resolved structure by BLINC and the excess level of 250 +/- 50 mJy. Constraining from the excess range of 5.8-10.5 $\micron$ region (IRAC 5.8 and 8 $\micron$, BLINC 10.5 $\micron$), a temperature of ~600 K is imposed for the inner hot component,  resulting in an excess peaking at ~5 $\micron$. In comparison, the hot component in our model has a lower temperature due to a different size distribution and grain composition which means that it matches the Stock et al. (2010) model well at >5 $\micron$ but was not fitted in the IRAC bands due to differences in the photospheric model. There is a difference of ~5\% in the photospheric fit used in the IRAC bands, but the photospheric subtraction at 10 and 24 $\micron$ produces values that are within the errors of those used by Stock et al. (2010), despite the difference in the stellar fit. When constraining the cold and warm components, we primarily consider the excess as >5 $\micron$ and refer to the Stock et al. (2010) model for a more thorough treatment of the hot excess.

The total mass in the collisional cascade, with the assumed size distribution
of $n(D)\propto D^{-3.5}$,  scales with $M_{tot}\propto\sqrt{D_{max}}$.
Thus scaling to the maximum grain size of 1 cm gives a dust mass of 2.1$\times$10$^{-4}$ M$_{\oplus}$, with 3.2$\times10^{-8}$ M$_{\oplus}$  of this in the hot component and the rest in the cold component.  Assuming the collisional cascade extends up to bodies of $D_{max}$=1 km then the total mass would be 1.2 M$_{\oplus}$. PSF variation can have an important effect on the observed residuals, so we repeated the modelling process using the Vesta PSFs and found no significant differences in the best fit model within the quoted errors, indicating that PSF variation has a negligible impact on the results.
 
\subsubsection{Comparison with  MCFOST Results}

We also compared the results of out model to the best fit found by the MCFOST code (\citealp{2006AA...459..797P}), a Monte Carlo radiation transfer code in which a star radiates isotropically in space and illuminates an azimuthally symmetric parametrised disc. This was used to fit both the Herschel 100 and 160 $\micron$ images and the entire SED of $\beta$ Leo. A grid of approximately a million models was run with the range of parameters described above and the joint $\chi^{2}$ of the images and SED was minimised. The best fit model derived from fitting the SED and PACS images with  MCFOST gives qualitatively similar best fit parameters as those derived from our IDL modelling suite (see $\S$ 3.2).  As the two approaches give similar results, this  validates the models within the constraints of the assumptions made about spatial and size distributions and compositions. The SED was weighted more heavily in calculating the $\chi^{2}$  in the MCFOST models, and the tension between the best-fit for the SED and images results in a compromise with the model preferring either most of the mass in a relatively narrow ring around 30-40 AU (as indicated by the SED), or a more extended disc starting further in (with smaller grains) and extending further out with a shallower surface density profile, which provides a better fit to the PACS images. As our model grid weighted the goodness of fit to the image surface brightness profiles more heavily than the SED fit, we consider the more extended disc as our best fit model. The 24 $\micron$ image was not included in the MCFOST grid, and the extension seen in this image provides evidence against the narrow ring interpretation.

\subsection{Three Component Model}
\label{3_comp}

In $\S$ \ref{best fit} we describe a 2 component model of the  $\beta$ Leo debris disc system. However, there is some compromise between the best fit models indicated by the SED and PACS images which leads to an uncertainty in the location of the inner edge of the disc between 15-30 AU, as this region is within the PACS beam size. This ambiguity is due to limitations from the initial assumption of a disc with an inner radius, an outer radius and a continuous surface density profile. By considering the possibility of a more complex 3 component model we can place better constraints on the inner region of the disc.  By combining the SED and surface brightness limits from the mid-IR and scattered light images we can place limits on possible 3 component models for the system and constrain any warm dust population between the hot and cold components. This 3 component model has 7 parameters: radius and cross-sectional area of hot dust ( r$_{hot}$, $\sigma_{hot}$), radius and cross-sectional area of warm dust (r$_{warm}$, $\sigma_{warm}$), radius and cross-sectional area of cold dust (r$_{cold}$, $\sigma_{cold}$) and composition of the warm component. We fix the composition of the hot and cold components to that inferred in $\S$3.2, since we know that this results in an appropriate range of dust temperatures at a given distance to fit the PACS images. The parameters of hot dust are set by the limits from the 10$\mu$m interferometry - here we restrict the parameter space to r$_{hot}$ of 2, 5 or 8 AU. The parameters of the cold dust are set by the PACS imaging as the best fit cold disc described in $\S$ 3.2. 

The main contribution from the cold component to the emission at 10-70 $\micron$ comes from near the inner edge of the distribution. We tried inner radii of the cold component of 15 AU, 20 AU, 25 AU and 30 AU, all of which are consistent with the PACS imaging. The warm component can then be constrained through SED fitting (See Table \ref{data}), primarily considering the fit to the IRS and MIPS photometry, and then checked for consistency with the surface brightness limits from the 11.3 and 18.1  $\micron$ Gemini imaging ($\S$2.3) and with the 24$\micron$ MIPS image ($\S$2.5).

The constraints on the warm component can be examined by considering the percentage of the total 18 $\micron$ flux that is produced by each component in the disc. If we assume the hot component is at 2 AU, then for this to fit the BLINC observations at 10 $\micron$, the warm component needs to produce 19\% of the total 18 $\micron$ flux, (48 mJy).  The cold component with an inner edge at 15 AU produces 81\% of the 18 $\micron$ flux, which means that there is no need for a warm component. This is shown in the first panel of Figure \ref{Disc_Pic}. If the inner edge is moved to 20 AU then only 70\% of the 18 $\micron$ emission is produced by the cold component, meaning that warm emission would have to account for 10\% (25 mJy) of the 18 $\micron$ emission. If the inner edge is at 30 AU then the cold disc only accounts for 51\% (129 mJy) of the total 18 $\micron$ emission, leaving 29\% (74 mJy) that must come from the warm component. 

If we repeat this analysis with a hot component at 5  AU (which produces 29\% of the 18 $\micron$ emission) we find that the inner edge of the cold component cannot be at 15 AU because this produces too much 18 $\micron$ emission, so the inner edge must be >20 AU. For an inner edge at 25 AU, we need a flux of 23 mJy at 18 $\micron$ (9\% of the total 18 $\micron$ flux) from the warm component. Similarly for an inner edge at 30 AU we need a flux of 51 mJy (20\% of the total 18 $\micron$ emission). These possible configurations are shown in Figure \ref{Disc_Pic}.

Although the above arguments imply that there are many possible hot dust radii and cold component inner edges, there are further constraints on the spectrum and many require unphysical assumptions about grain properties. There are two best fits to the 3 component model. The first has dust at 2 AU, 9 AU (Temperature of 160K) and 30-70 AU ($\gamma$=-2.0 for the cold component), whilst the second has dust at 5 AU, 12 AU and 30-70 AU ($\gamma$=-2.0). The same composition is used as that of the best fit 2 component model (See $\S$ \ref{composition}). The inteferometric constraints give slight preference to the first scenario with the dust at 9 AU. This gives us a $\chi_{r_{combined}}^{2}$=4.45 with $\chi_{rline_{100}}^{2}$=1.11, $\chi_{rline_{160}}^{2}$=1.31 and $\chi_{SED}^{2}$=2.03, showing that the 3 component model is a better fit than the 2 component model to the observations. As the 3 component model has more free parameters than the 2 component model (19 and 16 free parameters respectively) it is expected to have a lower $\chi^{2}$ than the 2 component model, but as the reduced $\chi^{2}$ takes into account the number of free parameters, $\nu$ they can be directly compared. The significance of this decrease in $\chi^{2}$ is discussed in $\S$ 3.4. The SED with this fit is shown in Figure 11.

\begin{figure*}
\caption{The SED fit for the 3 component model of the $\beta$ Leo disc. The SED shows the individual contributions (dotted lines) from a hot component at 2 AU, a warm component at 9 AU and a cold component at 30-70 AU. The stellar photosphere is shown as a straight line and the cumulative disc distribution is a solid black line. \label{3_comp_SED}}
\includegraphics[width=12.3cm]{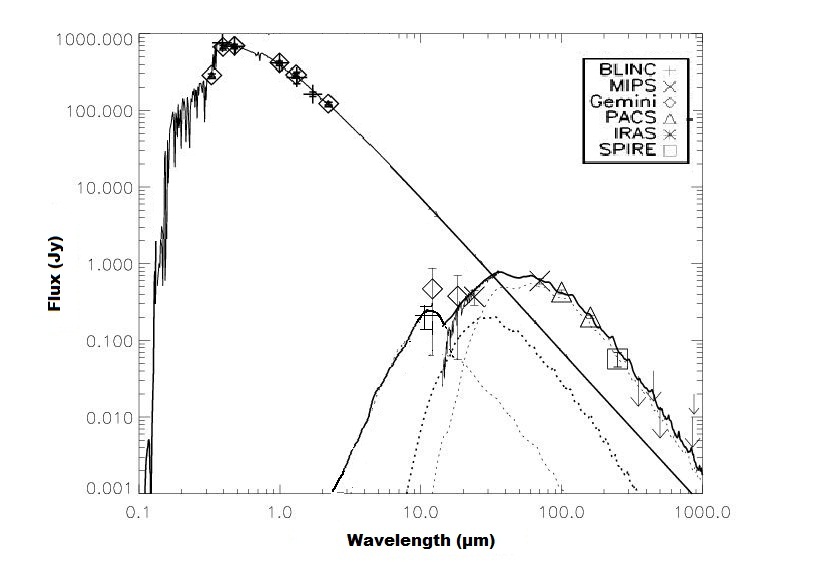}

\end{figure*}

\subsection{Eccentric Ring Model}
\label{ering}

Two other stars that harbour hot dust at $\sim$1 AU resolved interferometrically are HD69830 (\citealp{2006ApJ...639.1166B,2009A&A...503..265S}), and $\eta$ Corvi (\citealp{2009AA...503..265S}). Both these systems can be fit with a single continuous planetesimal population, in the form of a very eccentric (e>0.9) ring (\citealp{2010MNRAS.402..657W}) with a pericentre at the location of the hot dust and an apocentre corresponding to the cold belt for $\eta$ Corvi, and an as yet unseen cold population for HD69830. We used this model (\citealp{2010MNRAS.402..657W}) as a third option to try to explain the emission seen around $\beta$ Leo. This model assumes that the star formed with an eccentric planetesimal population and that this population has been evolving due to steady state collisional erosion for the 45 Myr age of the system. The model has two dust components: cold dust coincident with the planetesimals and hot dust created in collisions at pericentre that is being removed by radiation pressure. The model parameters are the pericentre, apocentre, maximum planetesimal diameter and the size of the grains being removed by radiation pressure. The best fit model was found by minimising the reduced $\chi^{2}$ fit to the SED and the 12, 18, 100 and 160 $\micron$ images across a grid of model parameters. The range of parameters tested to find the best fit were: pericentre from 0.5-5 AU (interval 0.5 AU), apocentre from 40-140 AU (interval 10 AU) and maximum planetesimal diameter between 500-3000 km (interval 500 km). The population of grains being removed by radiation pressure were assumed to be a single size, and values from 20-100 \% of the blow out size were tested. 

\begin{figure*} 
\caption{The SED of $\beta$ Leo modelled using a single planetesimal population with pericentre at 2 AU and apocentre at 65 AU. The dashed, dotted and solid lines correspond to the contribution from the collisional cascade, blow out grains and total emission spectrum respectively. The diagonal solid line is the stellar spectrum, the Spitzer IRS spectrum from 10-40 $\micron$ is plotted with a solid line, the asterisks are the MIPS points, the triangles photometry from Gemini Michelle and the squares PACS photometry. These are all shown with photospheric subtraction and have been colour corrected where necessary. \label{ering_sed}}
\includegraphics[width=12cm]{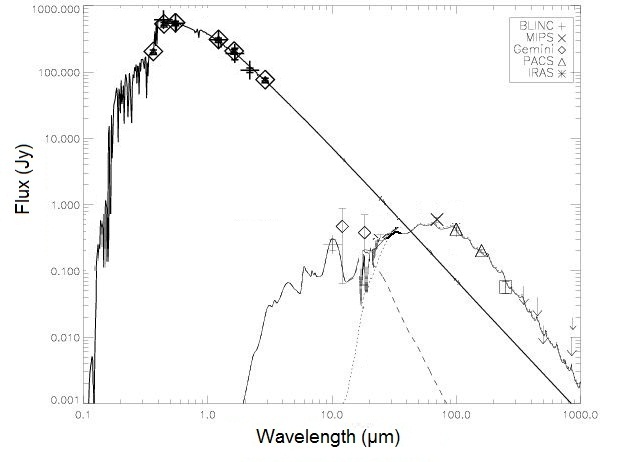}
\end{figure*}

\begin{figure*}
\caption{Surface brightness profiles for the $\beta$ Leo disc at 12 $\micron$ (top left), 18 $\micron$ (top centre), 24 $\micron$ (top right), 100 $\micron$ (bottom left) and 160 $\micron$ (bottom right) showing the fit of all 3 models to the observations. These are line cuts taken parallel to the major axis of the disc (125$^{\circ}$), summed over a width of 4 pixels. The observed profile is shown with a solid line, the model profiles with coloured dashed lines (red for the 2 component model, blue for the 3 component model and green for the eccentric ring model). The bottom panel of each plot show the residuals i.e. the observed linecut minus the model linecut to show the goodness of fit of the model. \label{ering_linecuts}}
\includegraphics[width=5.5cm]{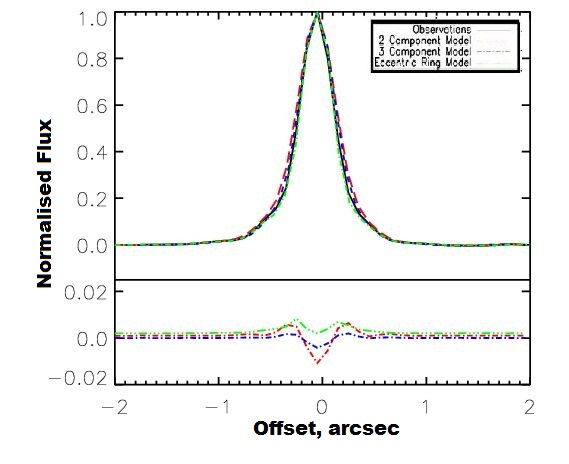}\hspace{0.25cm}\includegraphics[width=5.25cm]{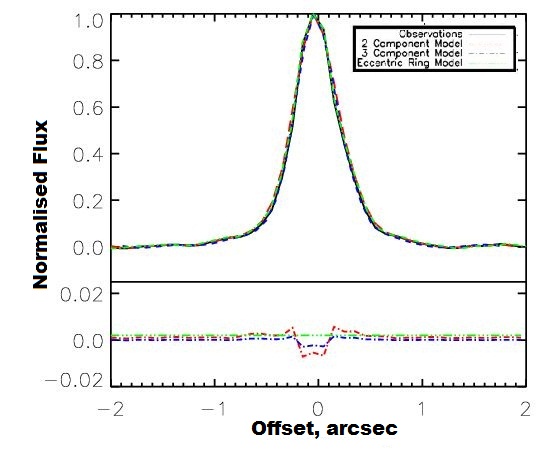}\hspace{0.25cm}\includegraphics[width=5.5cm]{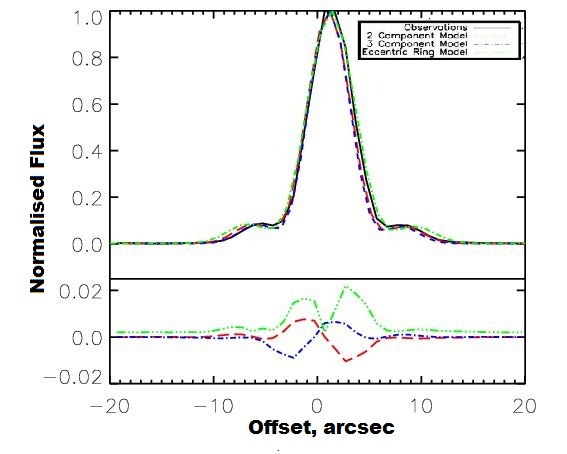}
\centering\includegraphics[width=5.5cm]{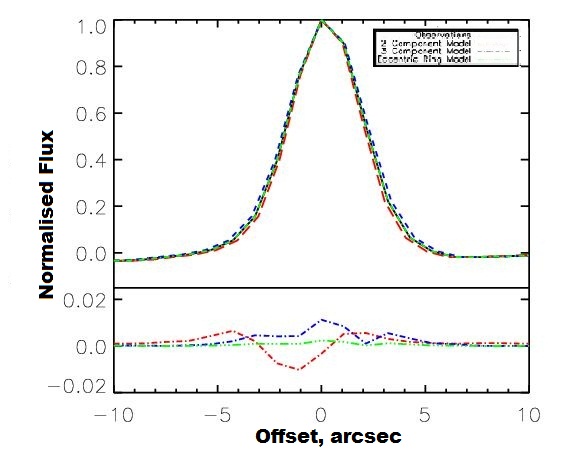}\hspace{0.25cm}\includegraphics[width=5.5cm]{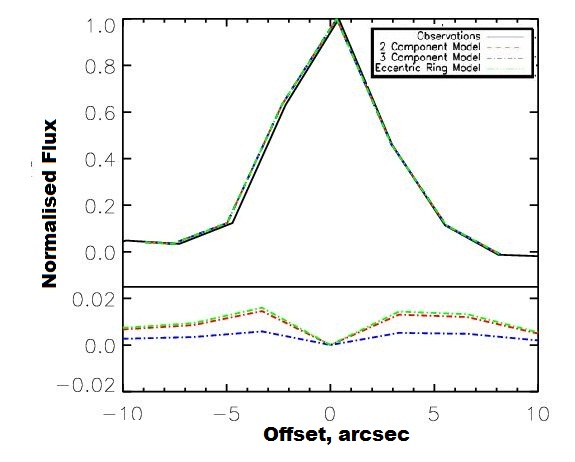}
\end{figure*}

 The SED for the best fit model is shown in Figure \ref{ering_sed}. This model has a single planetesimal population with pericentres at 2 AU and apocentres at 65 AU.  The eccentricity was therefore 0.96 and the maximum planetesimal diameter was 2000 km . The grains blown out from pericentre have a size of 2 $\micron$. The resulting mass-loss rate given by this model is 0.005 M$_{\oplus}$/Myr, giving a mass of 0.45 M$_{\oplus}$ for the parent planetesimal population. It is therefore possible to explain the $\beta$ Leo disc using one continuous planetesimal population, but the possible origins of such an eccentric planetesimal population are an issue (\citealp{2010MNRAS.402..657W}). The combined reduced $\chi^{2}$ for this model is $\chi_{r_{combined}}^{2}$=4.47, showing that the eccentric ring model has a slightly lower $\chi_{r_{combined}}^{2}$ than the 2 component model, but is a slightly worse fit than the 3 component model. The significance of this change in reduced $\chi^{2}$ is assessed in $\S$3.4. The fit to the surface brightness profiles of the 2 component, 3 component and eccentric ring models at 12, 18, 100 and 160 $\micron$  are shown in Figure \ref{ering_linecuts}. The 24 $\micron$ linecut shows that the eccentric ring model appears marginally more extended when compared with the observations. The FWHM of the eccentric ring model at this wavelength is  7.$\arcsec$0 x 6.$\arcsec$7 compared to a FWHM of  6."88 x 6."61 for the $\beta$ Leo observations. This means that the eccentric ring model is slightly less favoured, but cannot be ruled out.

\subsection{Ambiguity of the Models}

This addition of an extra parameters in the 3 component and eccentric ring models improves the $\chi^{2}_{r_{combined}}$ from 4.50 to 4.45 and 4.47 respectively, where a perfect fit to the resolved images at 100 and 160 $\micron$ and the SED would have  $\chi^{2}_{r_{combined}}$ =3.0. However, in general adding
an extra parameter will improve the fit of any model, so is adding this
extra parameter justified? Although the reduced $\chi^{2}$ compensates somewhat for this as it is divided by the number of free parameters, $\nu$, when considering models with large numbers of free parameters (19 in the case of the 3 component case) we need to avoid the problem of ``overfitting" or choosing a more complex model than is warranted by the data. We assess this using the Bayesian Information
Criterion (BIC):
\begin{equation}
BIC = Nln(\chi^{2}_{r_{combined}} ) + klnN
\end{equation}
where N is the number of data points, k is the number of
free parameters and $\chi^{2}_{r_{combined}}$  is the minimum combined reduced
$\chi^{2}$ for the model. The BIC considers the fit of the model but penalises the model for extra parameters (see \citealp{2005ApJ...618..385W, 2007MNRAS.377L..74L}). It penalises the model more strongly for extra parameters than the reduced $\chi^{2}$ and so provides a more stringent test of the validity of applying more complex models. The BIC value for the 2 component model is 51.1, for the
3 component model it is 49.7, and for the eccentric ring model the BIC is 51.9. A lower value of the BIC is preferred; a difference of 2
between BICs of different models indicates positive evidence
against the higher BIC value, and a difference of 6 indicates
strong evidence against the higher BIC value. Therefore our results indicate that there is no real preference between the 3 models since they all reproduce the data well. Furthermore it is not unreasonable to assume that a realistic disc system may well be more complicated that the simple parameterisations used in these models.

\begin{figure*}
\caption{Predicted 11.3 $\micron$ surface brightness profiles for the three different models of the $\beta$ Leo disc: the 2 component model with a hot disc at 2-3 AU and a cold disc from 15-70 AU; a 3 component model with a hot disc at 2-3 AU; a warm ring at 9 AU and a cold disc from 30-70 AU and the eccentric ring model\label{IR_sb}.}
\includegraphics[width=8cm]{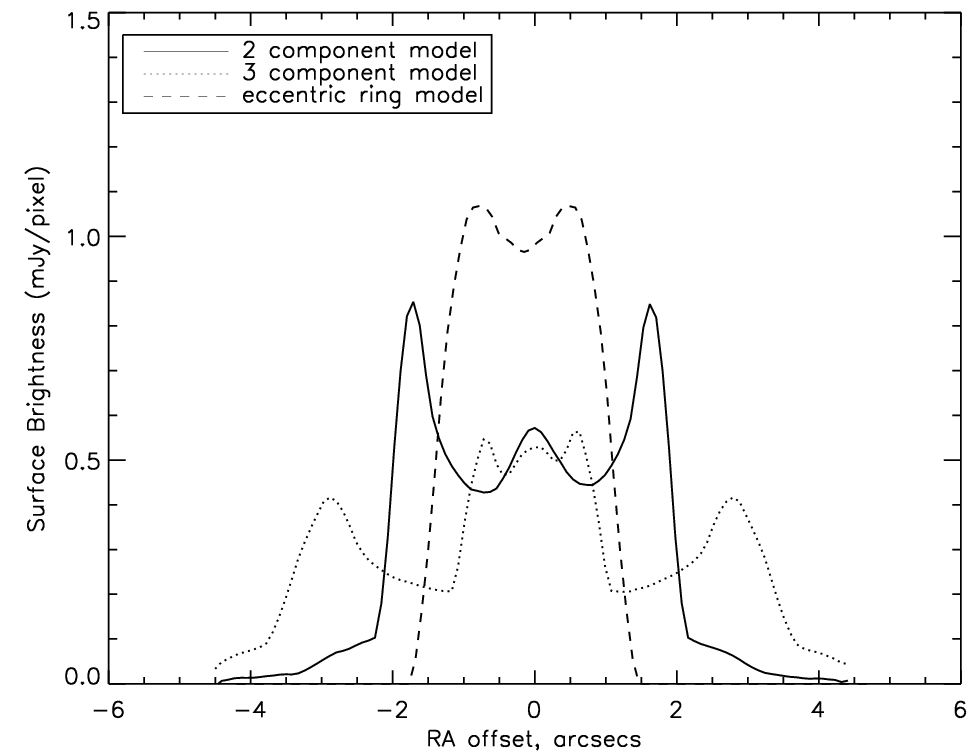}
\end{figure*}

\begin{figure*}
\caption{An illustration of the possible structures of the $\beta$ Leo disc based on the 3 different possible models of the disc. The top panel shows the 2 component model, which has belt at 2 AU, a gap from 2-15 AU  and an extended disc from 15-70 AU. The middle panel shows the 3 component model which has a belts at 2 AU, 9 AU and 30-70 AU. The bottom panel shows the eccentric ring model as described in the text. \label{Disc_Pic}}
\includegraphics[width=11.0cm]{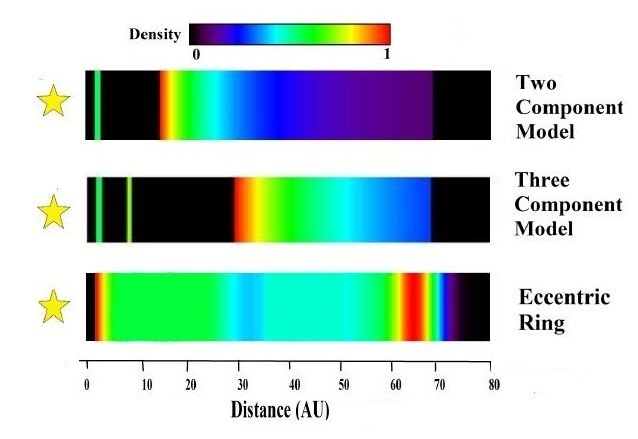}
\end{figure*}
These three models all provide a very good fit to the observations, but further data could allow us to resolve this ambiguity. The best way to distinguish between the possible 3 component models and the 2 component model would be to resolve the inner edge or the warm component. This may be possible with very deep mid-IR imaging in the N band. Resolving the disc at this wavelength would give different surface brightness profiles for the different models as shown in Figure \ref{IR_sb}. The Spitzer image at 24 $\micron$ already suggests that the extent of the eccentric ring model may be too large. A quadratic subtraction of the FWHM of the beam from the image FWHM diameter (after PSF subtraction) indicates a diameter of $\sim$ 4 $\arcsec$ FWHM for the disc at this wavelength, significantly less than the value indicated in Figure 13.

It  may also be possible to resolve the inner edge of the cold disc in scattered light. We created simulated ACS images for all three models to compare to $\S$2.6. Mie theory was used to calculate the albedo of grains in the model and a bandpass of 0.45-0.72  $\micron$  was assumed for the F606w filter.  A Henyey-Greenstein phase function (\citealp{1941ApJ....93...70H}) was used to approximate the asymmetric scattering by small particles with an asymmetry parameter (g) of 0.3 corresponding to slightly forward scattering. Due to uncertainties in scattering properties we might expect our predicted surface brightnesses to have significant uncertainties. However, the morphology for the emission (and relative level of surface brightness for different models) should be correct.

The ACS coronograph has an inner working angle of 1.8 $\arcsec$ so in all the models the hot component is hidden behind the occulting spot. For the 2 component model the peak surface brightness in scattered light would be a radius of 1.35 $\arcsec$ (i.e. behind the occulting spot) with a surface brightness of 1.7$\times$10$^{-3}$  mJy/pixel, assuming a pixel size of 25 mas. Just outside the inner working angle of the coronograph (1.9 $\arcsec$ radius) the disc would have a surface brightness of 1.3$\times$10$^{-3}$  mJy/pixel and fall off $\propto$ r$^{-3.5}$. The three component model has peak surface brightness of 2.1$\times$10$^{-4}$ mJy/pixel at a radius of 2.7 $\arcsec$ and the eccentric ring model has a peak surface brightness of 4.6$\times$10$^{-4}$ at a radius of 3.6\arcsec. In an annulus $2.8"-3.0"$ the residual halo in the HST ACS observations shown in $\S$\ref{ACS_obs} has a median brightness $5.3\times10^{-4}$ mJy/pixel, which is consistent with the three component model.  The halo brightness decreases steeply to the sensitivity-limited sky value at $4.0 \arcsec$ radius.  If the halo were due to grain scattering, we would expect a radial brightness profile proportional to $r^{-2}$ if the star was embedded in a uniform density sheet of dust ($\gamma$=0).  The measured radial brightness profile is proportional to $r^{-6}$, implying that $\gamma$$\sim$-3.  This is comparable to the dust radial profile for the southwest midplane of the $\beta$ Pictoris dust disc (\citealp{1995AJ....110..794K, 2006AJ....131.3109G}).  However, this is significantly steeper than $\gamma$=-2.0 for the three component model.

The nature of this residual emission is unclear - it is difficult to find a disc model that can both produce sufficient scattering in this region without producing a very bad fit to the SED. Models with an outer edge at 44 AU, as seen in the ACS observations or with $\gamma$<-2.0 to mimic the steep drop-off seen are also a very bad fit to the extended emission seen with PACS at 100 $\micron$ and 160 $\micron$. However, it could be reconciled with the two component model if we are seeing a separate small grain population in the optical observations that has a steeper surface density profile.  If this emission was real, then it could be used to place limits on the albedo of the grains in the eccentric ring model, or possibly rule out this scenario. Currently, the observations favour the two component model as no peak is seen at 2.7 or 3.6\arcsec. Further optical imaging could confirm the nature and location of this emission and possibly rule out the eccentric ring model, which has the largest surface brightness in this region.

\section{Discussion}

The $\beta$ Leo debris disc system consists of multiple dust populations. We can find a best fit solution with either 1, 2 or 3 components. These are shown in Figure \ref{Disc_Pic}. The 2 component model is the simplest and has dust at 2-3 AU and 15-70 AU. There are two best fits to the 3 component model. The first has dust at 2 AU, 9 AU and 30-70 AU, whilst the second has dust at 5 AU, 12 AU and 30-70 AU. The inteferometric constraints give slight preference to the first scenario. The key constraint on these models is that the warm (middle) component cannot be located inside an 8 AU radius without violating the 10 $\micron$   interferometry constraints.  A third possible scenario for this system is that the emission results from a single very eccentric (0.96) planetesimal population.  This model consists of a steady-state collisional cascade with an apocentre coinciding with the location of the cold emission seen in the PACS images, with an additional population of hot, small grains created at collisions near pericentre which are subsequently removed from the system due to radiation pressure.

These inferred structures give us clues to the underlying dynamics of the system, as the belts are confined and separated by gaps suggesting dynamical interactions. The 2 component model has a gap between 2 and 15 AU which could contain planets, as could the gaps between 2 and 9 AU and between 10 and 30 AU in the 3 component model.

\subsection{Origin of the Hot Dust}
The origin of the hot, compact dust populations is still unclear.
There are five possible scenarios: i) the small dust grains produced by collisions in
the cold belt could drift towards the inner region due to PR drag; ii) the hot dust could be evidence of a  planetesimal population in a steady state collisional cascade at a few AU from the star; iii) the dust could be transient,  produced in a massive collision between planetesimals;  iv) the system could be undergoing a dynamical instability resulting in the planetesimals being thrown in from the outer belt and producing hot dust as in the Late Heavy Bombardment (\citealp{2005Natur.435..466G}); or v) the grains may be produced through
evaporation of comets originating from the cold planetesimal belt
imaged in the sub-mm, like the Zodiacal Cloud (\citealp{2010ApJ...713..816N}).

The first scenario concerning PR drag as an origin for the hot dust is improbable due to the long timescale for PR drag compared to the shorter
collisional timescale for the outer disc (\citealp{2007ApJ...663..365W}).
This mechanism is  unlikely to produce the amount of dust
observed in $\beta$ Leo's inner regions. Around late type stars radiation pressure is inefficient at removing small grains, so these can be transported into the inner regions of the disc via Poynting-Robertson drag and stellar wind drag, even in discs with optical depths considered too high for discs to be transport-dominated  ($\tau$ $\sim$10$^{-4}$ \citealp{2010arXiv1011.4882R}). This is because close to the star larger grains are preferentially collisionally eliminated, causing a break in the size distribution at a critical size, below which grains are transport dominated. Since $\beta$ Leo is an A type star, however, the larger radiation pressure blow-out size will mean that all the grains remaining in the disc are larger than the critical size and hence are collisionally-dominated. 

 The feasibility of a steady-state belt as an explanation for the hot dust can be examined by considering the maximum steady state fractional luminosity of the belt. The fractional luminosity of the hot component from our model is 8.1$\times$10$^{-5}$. Using equation 18  from \citet{2008ARAA..46..339W} for the maximum fractional luminosity of a belt at a given age;

\[
 f_{max}=0.58\times10^{-9}r^{7/3}(dr/r)D_{c}^{0.5}Q^{*5/6}_{D}e^{-5/3}M_{*}^{-5/6}L_{*}^{-0.5}t_{age}^{-1}
\]
\hspace{-0.1cm} and assuming a radius (r) for the hot component of 2.5 AU, a width (dr) of 1 AU, a maximum planetesimal diameter ($D_{c}$) of 60 km, planetesimal strength  (Q$_{d}^{*}$) of 150 J/kg (the fiducial value from \citet{2008ARAA..46..339W}),  an eccentricity (e) of 0.05, stellar mass (M$_{*}$) of 2.1 $M_{\odot}$, a luminosity of 14.0 L$_{\odot}$ and an age of 45 Myr, gives a maximum fractional luminosity of 9.9$\times$10$^{-7}$. For the hot dust population to be considered transient,  \citet{2007ApJ...663..365W}  concluded that the fractional luminosity must be 1000 $F_{max}$, so $\beta$ Leo's hot dust is consistent with a steady state model, but only marginally so, and to be a steady state phenomenon  would have to have properties that are significantly different from those for other A stars (e.g. unusually strong or large planetesimals). This implies that the third scenario is unlikely as again, a very large belt mass would be required due to the low probability of a massive collision having occurred and the short lifetime of the dust produced.

 The fourth scenario also interprets the hot dust as a transient phenomenon by
suggesting that the system is undergoing major dynamical perturbations
in a scenario analogous to the Late Heavy Bombardment (LHB) in the solar
system. The LHB occurred 700 Myr after the planets formed and led to a
spike in the production of dust  as comets were thrown in towards
the star by a dynamical instability caused by the migration of the
giant planets (\citealp{2005Natur.435..466G}). It is not known how
common such events are, although \citet{2009MNRAS.399..385B} estimate
that they occur around <12\% of sun-like
stars.  Since an LHB-like event could occur at any point in a disc's lifetime and as only a small percentage of stars undergo such an event, it is improbable that we are observing such a disturbance in the $\beta$ Leo disc.

This leaves the final scenario, that the grains may be produced through
disintegration of comets originating from the cold planetesimal belt
imaged in the sub-mm. Since this requires neither a very high initial mass for the disc nor a low-probability event such as a collision or a LHB like instability, it seems more likely although the dynamics of scattering and physics of dust production remain poorly understood. If the hot dust does originate from a population of scattered planetesimals then the amount of hot dust could be indicative of  the configuration of the planetary system responsible for the scattering e.g. the effect of Jupiter on scattering of planetesimals in the Solar System \citep{2008IJAsB...7..251H, 2009IJAsB...8...75H,2010IJAsB...9....1H}, This implies that there could be multiple planets between the dust belts responsible for scattering planetesimals into the inner regions of the system, and so replenishing the hot dust population.

\begin{table*}
\caption{Table comparing radius, age and fractional luminosity of debris discs resolved in thermal emission around A stars }

{\small }\begin{tabular}{|p{2cm}|p{1.05cm}|p{2cm}|p{0.8cm}|p{1cm}|p{1.5cm}|p{1.5cm}|p{1.5cm}|p{0.7cm}|p{2cm}|}
\hline 
{\small Name} & {\small Wavelength Resolved ($\micron$) } & {\small Spectral Type} & {\small Inner Radius (AU)} & {\small Outer Radius (AU)} & Fractional Luminosity (cold component) & {\small Hot Component} & {\small fractional luminosity (hot component)} & {\small Age (Myr)} & {\small Notes}\tabularnewline
\hline

{\small Fomalhaut (HD216596)} & {\small 1$ $$^{[1]}$, 450$ $$^{[2]}$, 850$ $$^{[3]}$} & {\small A3V$^{[1]}$} & {\small 135$^{[1]}$, 60$^{[2]}$,  60$^{[3]}$} & {\small 160$^{[1]}$, 150$^{[2]}$, 150$^{[3]}$} & {\small 4.6$\times$10$^{-5}$ $^{[4]}$} & {\small <6 AU$^{[4]}$} & {\small 5$\times$10$^{-4}$$^{[4]}$} & {\small 200$^{[5]}$} & {\small$<$3$M_{jup}$ Planet at$\sim$130
 AU$^{[7]}$}\tabularnewline
\hline 
{\small Vega (HD172167)} & {{\small 24$ $$^{[7]}$ 70$ $$^{[7,8]}$, 160$ $$^{[7,8]}$,
250$ $$^{[8]}$, 350$ $$^{[8]}$, 500$ $$^{[8]}$, 850$ $$^{[2]}$,
1.1mm$^{[9]}$, 3mm$^{[9]}$}} & {\small A0V$^{[4]}$} & {\small 86$^{[7]}$, $\sim$85$^{[8]}$, $\sim$85
$^{[8]}$, $\sim$85 $^{[8]}$, $\sim$85 $^{[8]}$,
80$^{[2]}$, 80$^{[9]}$, 80$^{[9]}$} & {\small 330$^{[7]}$, 543$^{[7]}$, 815$^{[7]}$, $\sim$800 $^{[8]}$,
 $\sim$800 $^{[8]}$, $\sim$800 $^{[8]}$, $\sim$800$^{[2]}$,
$\sim$800$^{[9]}$, $\sim$800$^{[9]}$ } & {\small 2.3$\times$10$^{-5}$ $^{[11]}$} & {\small <8 AU$^{[10]}$} & 5.1$\times$10$^{-4}${\small $^{[5]}$} & {\small 200$^{[11]}$} & {\small Halo of small grains$^{[7]}$}\tabularnewline
\hline 
{\small HR4796A (HD109573)} & {\small 1.2$ $$^{[12]}$, 10$ $$^{[13]}$, 18$ $$^{[13,14]}$,
24$ $$^{[14]}$} & {\small A0V$^{[12]}$} & {\small 70 $^{[15]}$} & {\small 84$^{[15]}$} & {\small 5$\times$10$^{-3}$ $^{[12]}$} & {\small N/A} & N/A & {\small 10$^{[12]}$} & radius from multi-wavelength model\tabularnewline
\hline 
{\small $\eta$ Tel (HD181296)} & {\small 18$ $$^{[16]}$} & {\small A1V$^{[16]}$} & {\small 21 $^{[16]}$} & {\small 26$^{[16]}$} & {\small 1.4$\times$10$^{-4}$ $^{[16]}$} & {\small <4 AU$^{[16]}$} & 1.6$\times$10$^{-4}$$^{[16]}$ & {\small 12$^{[17]}$} & \tabularnewline
\hline 
{\small $\zeta$ Lep (HD38678)} & {\small 18$ $$^{[18]}$} & {\small A2V$^{[11]}$ } & {\small 3 AU$^{[18]}$}&{\small 8 AU$^{[18]}$} & {\small 6.5$\times$10$^{-5}$$^{[11]}$} &{\small 0.2$^{[4]}$}&{\small 2.5$\times$10$^{-3}$} & {\small 230$^{[18]}$} & \tabularnewline
\hline 
{\small $\beta$ Pictoris (HD39060)} & {\small 0.5$ $$^{[19]}$, 1.2$ $$^{[20]}$, 10$ $$^{[21]}$,
850$ $$^{[2]}$} & {\small A5V$^{[11]}$} & {\small 50$^{[19]}$, 50$^{[20]}$, 25$^{[21]}$, 50$^{[2]}$} & 700$^{[19]}$, >100$^{[20]}$ >100$^{[21]}$, 250$^{[2]}$ & {\small 3$\times$10$^{-3}$$^{[11]}$} & {\small 6 $^{[44]}$} & {\small 2$\times$10$^{-3}$$^[44]$} & {\small 12$^{[19]}$} & {\small 8 Jupiter Mass planet at 8-15 AU$^{[22]}$}\tabularnewline
\hline 
{\small $\beta$ Leo (HD102647)} & {\small 100$ $$^{[23]}$ 160$ $$^{[23]}$} & {\small A3V$^{[24]}$} & {\small 15$^{[25]}$ } & 70$^{[25]}$ & {\small 3$\times$10$^{-5}$ $^{[27]}$} & {\small 2-3 AU$^{[26]}$} & 7.8$\times$10$^{-5}$ $^{[26]}$ & {\small 45$^{[28]}$} & \tabularnewline
\hline 
{\small $\gamma$ Oph (HD161868)} & {\small 24$ $$^{[28]}$, 70$ $$^{[28]}$} & {\small A0V$^{[29]}$} & {\small 13$^{[28]}$} &  430$^{[28]}$ & {\small 9$\times$10$^{-5}$$^{[28]}$} & {\small N/A} & {\small N/A} & {\small 180$^{[30]}$} & \tabularnewline
\hline 
{\small HD141569} & {\small 0.5$ $$^{[31]}$, 1.1$ $$^{[32]}$, 1.6$ $$^{[32]}$,
12.5$ $$^{[33]}$, 18$ $$^{[33]}$} & {\small A0V$^{[11]}$} & {\small 175$^{[31]}$, 190$^{[32]}$, 30$^{[33]}$} & {\small 1200$^{[31]}$, 360$^{[32]}$, 150$^{[33]}$} & {\small 4.6$\times$10$^{-3}$$^{[11]}$} & {\small N/A} & {\small N/A} & {\small 5$^{[11]}$} & {\small Spiral Structure; Possibly due to giant planets$^{[35]}$}\tabularnewline
\hline 
{\small 49 Ceti (HD9672)} & {\small 12.5$ $$^{[36]}$, 18.0$ $$^{[36]}$} & {\small A1V$^{[11]}$} & {\small 30 $^{[36]}$} & {\small 60$^{[36]}$} & {\small 7.2$\times$10$^{-4}$ $^{[11]}$} & {\small N/A} & {\small N/A} & {\small 20$^{[11]}$} & \tabularnewline
\hline 
{\small HD32297} & {\small 0.8$ $$^{[37]}$, 1.1$ $$^{[37]}$, 1.6$ $$^{[38]}$,
2.02$ $$^{[38]}$, 12$ $$^{[39]}$, 18$ $$^{[39]}$} & {\small A0V$^{[11]}$} & {\small 560$^{[37]}$, 50$^{[38]}$, 80$^{[38]}$} & {\small 1680$^{[37]}$, 400$^{[38]}$, 300$^{[39]}$} & {\small 2.7$\times$10$^{-3}$$^{[37]}$} & {\small N/A} & {\small N/A} & {\small 30$^{[37]}$} & {\small Sculpted by the ISM?$^{[39]}$} Warped Disc$^{[37,39]}$\tabularnewline
\hline 
{\small HR 8799 (HD218396)} & {\small 24$ $$^{[41]}$, 70$ $$^{[41]}$, 160$ $$^{[41]}$} & {\small A5V$^{[41]}$} & {\small 90$^{[41]}$} & {\small 300$^{[41]}$} & {\small 4.9$\times$10$^{-5}$ $^{[11]}$} & {\small 6-15$^{[41]}$} & {\small N/A} & {\small 20-160$^{[41]}$} & {\small Halo of small grains out to >1000 AU$^{[41]}$}\tabularnewline
\hline
$\beta$ UMa (HD95418) & 11.2$ $$^{[42]}$, 18.1$ $$^{[42]}$, 100$ $$^{[23]}$ & A1V & 50 & 50 & 1.4$\times$10$^{-5}$  & 1.09$\pm$0.01 $^{[42]}$ & ~10$^{-5}$$^{[42]}$ & 50$^{[30]}$, 520$^{[44]}$ & \tabularnewline
\hline
HD139006 & 11.2$ $$^{[42]}$, 18.1$ $$^{[42]}$ & A0V & 46$^{[43]}$  & 46$^{[43]}$ &  1.2$\times$10$^{-5}$ $^{[43]}$ & 2.31$\pm$0.04 $^{[42]}$  & 1$\times$10$^{-5}$$^{[42]}$ & 314$^{[30]}$, 350$^{[44]}$ & Eclipsing Binary (companion G5V, 0.9M$_{\odot}$ at 0.2  AU)  \tabularnewline
\hline
HD181869 & 10.4$ $$^{[42]}$, 18.1$ $$^{[42]}$ & B8V &   N/A & N/A &  N/A& 4.12$\pm$0.11 $^{[42]}$  & 1$\times$10$^{-5}$$^{[42]}$ & 100$^{[44]}$ & \tabularnewline
\hline
HD3003 & N/A & A0V$^{[43]}$ & 14$^{[41]}$ & 24$^{[41]}$ & 2.10$\times$10$^{-4}$ $^{[41]}$ & 4-6.5$^{[43]}$ & 2.01$\times$10$^{-4}${\small $^{[43]}$} & {\small 50$^{[30]}$} & \tabularnewline
\hline 
\end{tabular}
\end{table*}
\begin{table*}
{\small }\begin{tabular}{|p{2cm}|p{1.05cm}|p{2cm}|p{0.8cm}|p{1cm}|p{1.5cm}|p{1.5cm}|p{1.5cm}|p{0.7cm}|p{2cm}|}
\hline 
{\small Name} & {\small Wavelength Resolved ($\micron$) } & {\small Spectral Type} & {\small Inner Radius (AU)} & {\small Outer Radius (AU)} & Fractional Luminosity (cold component) & {\small Hot Component} & {\small fractional luminosity (hot component)} & {\small Age (Myr)} & {\small Notes}\tabularnewline
\hline

\hline
HD23281 & N/A & A6V$^{[43]}$ & N/A & N/A & N/A & 5.4-9.5$^{[43]}$ & 3.82$\times$10$^{-5}$ $^{[43]}$ & {\small 626$^{[30]}$} & \tabularnewline
\hline 
$\lambda$ Gem (HD56537) & N/A & A3V$^{[43]}$ & N/A & N/A & N/A & 2.1-6.1$^{[43]}$ & 5.64$\times$10$^{-5}$$^{[43]}$ & {\small 560$^{[43]}$} & \tabularnewline
\hline 
HD71155 & N/A & A0V$^{[43]}$ & 44$^{[11]}$ & 90$^{[43]}$ & 2.5$\times$10$^{-5}$$^{[11]}$ & 2.2-8.2$^{[43]}$ & 8.95$\times$10$^{-5}$ $^{[43]}$ & {\small 169$^{[30]}$} & \tabularnewline
\hline 
HD80950 & N/A & A0V$^{[43]}$ & 13.6$^{[43]}$ & 24.0$^{[43]}$ & 9.62$\times$10$^{-5}$ $^{[43]}$ & N/A & N/A & {\small 80$^{[45]}$} & \tabularnewline
\hline 
HD141795 & N/A & A2m$^{[43]}$ & N/A & N/A & N/A & 4.6-6.1$^{[43]}$ & 4.43$\times$10$^{-5}$$^{[43]}$ & {\small 450$^{[30]}$} & \tabularnewline
\hline
\end{tabular}{\small \par}

\vspace{0.2cm}

{\small { \bf References}: 1)\citet{2005Natur.435.1067K}, 2)\citet{1998Natur.392..788H}
3)\citet{2003ApJ...582.1141H} 4)\citet{2009ApJ...704..150A} 5)\citet{2004AA...426..601D}
6)\citet{2008Sci...322.1345K} 7)\citet{2005ApJ...628..487S} 8)\citet{2010AA...518L.130S}
9)\citet{2002ApJ...569L.115W} 10)\citet{2006AA...452..237A} 11)\citet{2007ApJ...663..365W}
12)\citet{1999ApJ...513L.127S} 13)\citet{2000ApJ...530..329T} 14)\citet{2010ApJ...530..329T}
15) \citealp{1999ApJ...527..918W} 16)\citet{2009AA...493..299S}
17)\citet{2009AJ....137.3632L} 18)\citet{2007ApJ...655L.109M} 19)\citet{2000ApJ...530L.133K}
20)\citet{1994Natur.369..628L} 21)\citet{1997MNRAS.292..896M} 22)\citet{2010Sci...329...57L}
23)\citet{2010AA...518L.135M} 24)\citet{2010MNRAS.403.1089P} 25)
This Work 26) Stock et al. (2010) 27)\citet{2006ApJS..166..351C} 28)\citet{1999AA...348..897L}
29)\citet{2008ApJ...679L.125S} 30)\citet{2001ApJ...546..352S} 31)\citet{2003AJ....126..385C}
32)\citet{1999ApJ...525L..53W} 33)\citet{1998AAS...193.7316S} 35)\citet{2005AA...440..937W}
36)\citet{2007ApJ...661..368W} 37)\citet{2005ApJ...635L.169K} 38)\citet{2005ApJ...629L.117S}
39)\citet{2007ApJ...666L.109M} 40)\citet{2009ApJ...702..318D} 41)\citet{2009ApJ...705..314S}
42)\citet{2010ApJ...723.1418M}43)\citet{2010AA...515A..95S} 44)\citet{2005ApJ...620.1010R}
45)\citet{2003ApJ...584L..27W}}

\end{table*}

\subsection{Placing $\beta$ Leo into Context}

There are currently 16 debris discs around A type stars that have at least one component resolved in thermal emission.  These are listed in Table 3. Of these 16 discs 11 show evidence of multiple dust components, where at least one component has been imaged. The last 5 objects in Table 3 are A stars with evidence of multiple dust components including hot dust, but whose discs have not been resolved. Of these A stars, most are thought to be younger than $\beta$ Leo, with ages
in the range of 10s of Myr.  

The discs  summarized in Table 3 show a startling diversity, with radii from a few AU, i.e., $\zeta$ Lep (3-8 AU \citealp{2007ApJ...655L.109M}) to hundreds of AU, i.e., Vega (\citealp{2005ApJ...628..487S} ). Some systems are confined rings with clear inner and outer edges, i.e., HR4796 (\citealp{1999ApJ...513L.127S})  and some have very extended discs like $\beta$ Pic (\citealp{2000ApJ...530L.133K}).

\begin{figure*} 
\caption{Plot showing the radius and fractional luminosity of the hot and cold components for the A star debris discs listed in Table 3 as having 2 components.\label{A_stars}}
\includegraphics[width=15cm]{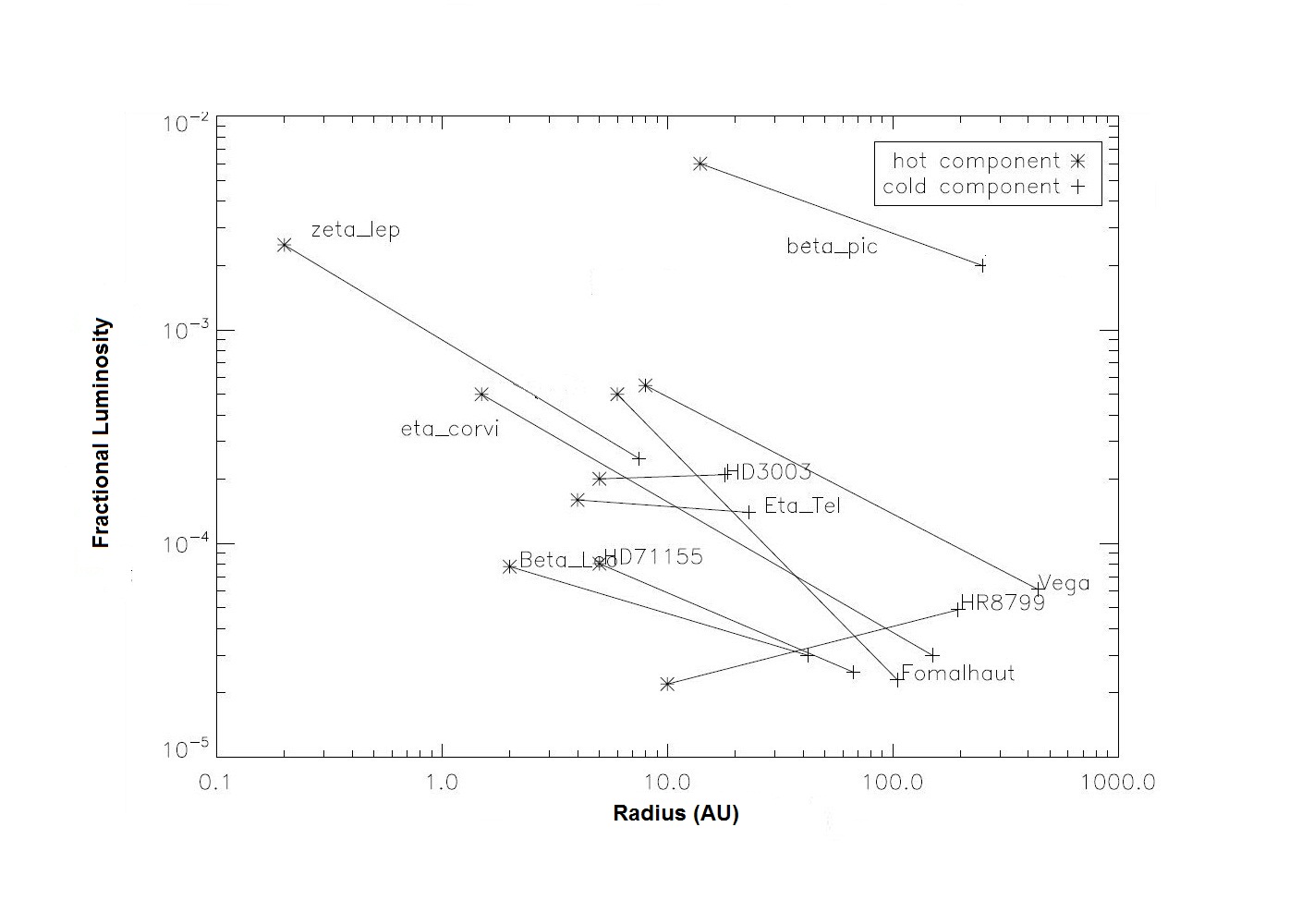}
\end{figure*}

 Figure \ref{A_stars} shows a plot of the fractional luminosity vs radius of  both the hot and cold disc component for all the A stars listed in Table 3 as having  2 disc components. Vega, Fomalhaut and $\beta$ Leo are the only A stars with a hot disc component that has been resolved via  interferometry (\citealp{2006AA...452..237A, 2009ApJ...704..150A,2009ApJ...691.1896A}). The hot dust in other discs is either inferred from SED fitting (as for HR4796 and HR8799) or resolved in the mid-IR ($\zeta$ Lep, $\beta$ UMa, HD139006). This leads to potential systematic differences when comparing radii and fractional luminosities derived through different methods. These warm components could also be related to planetesimals near the ice-line (\citealp{2011ApJ...730L..29M}). Although there are no clear trends in the small number of objects with known 2 component discs, there are some interesting points to note. 

We see two main 'types' of distributions between the hot and cold populations. First, discs such as  Vega, Fomalhaut, $\beta$ Leo and $\eta$ Corvi, which have a large (>20 AU) separation between the location of the hot and cold dust, and whose hot dust has a high 
fractional luminosity. This leads to a steep line in Figure \ref{A_stars}. The other type of disc, typified by Eta Tel, HD3003 and HR8799 show an almost flat or, for HR8799, declining line in Figure \ref{A_stars}, indicating that the hot dust has a similar or lower fractional luminosity than the cold component. In the case of HR8799 the hot and cold components are separated by a known planetary system  (\citealp{2008Sci...322.1348M}) which may inhibit the transfer of material from the outer to inner regions of the system. 

Fomalhaut, Vega and $\beta$ Pic show the steepest increase in fractional luminosity between cold and hot components. In all 3 of these cases the hot component is derived from modelling fitting the interferometric visibility deficit seen around these objects (\citealp{2006AA...452..237A,2009ApJ...704..150A,2010arXiv1009.1245A}). $\beta$ Leo shows a gradient between the hot and cold components, similar to that of $\eta$ Corvi, an F star with multiple dust populations including a cold dust disc resolved with PACS and SCUBA at  150 AU (\citealp{2010AA...518L.135M}) and $\zeta$ Lep, which has a much more compact configuration, with its 'cold' disc at only 10 AU.

 In the Solar System the asteroid belt (radius $\sim$2 AU, f=0.8$\times$10$^{-7}$ \citep{1993prpl.conf.1253B}) and the Kuiper Belt (radius$\sim$30 AU, f$\sim$10$^{-7}$ \citep{2005Natur.435..466G} would also give a flat line, similar to that of HR8799. The radial distribution of dust in these systems could be indicative of  the underlying planetary system. HR8799 and the Solar System both have multiple planet systems between their dust belts which could influence the levels of planetesimal scattering. However,  Fomalhaut and $\beta$ Pic have at least one planet \citep{2008Sci...322.1345K, 2010Sci...329...57L} but still have high hot dust levels and a steep line on Figure \ref{A_stars}. 

\subsection{Implications for Possible Planetary Systems}

This diversity in disc structure and brightness seen in these discs with hot and cold components could be due to the underlying dynamics of the system, which can be affected by the presence and distribution of planets in the system. For example,  in the Solar System the number of comets thrown into the inner system from the Kuiper belt is dependent on planetary architecture (\citealp{2010IJAsB...9....1H}). For the $\beta$ Leo disc, the radial distributions of dust that we have derived have implications for the architecture of possible planetary systems.  For the 2 component model there is a gap in the disc between 2 and 15 AU, in which there could be planets that are truncating the cold disc and confining the hot inner disc. For the 3 component model the gaps from 3-9 AU and 9-30 AU could occur again due to planets truncating the discs. The size of these gaps  could be linked to clearing by a planet due to the scattering of planetesimals out from the planet's chaotic zone. The area cleared increases for multiple planets and for planets on eccentric orbits (\citealp{2011arXiv1102.3185B}). In the case of multiple planet systems these could influence the amount of material scattered into the inner solar system and the possible stable locations for dust belts, so the morphology of the dust belts in the Beta Leo system could indicated multiple planets between the cold and hot dust belts.

\section{Conclusions}

We have presented detailed modelling of the debris disc around the 45 Myr old A star $\beta$ Leo. We considered multi-wavelength data  to construct a complete picture of this source. Resolved images taken at 100 and 160 $\micron$ using PACS on Herschel as part of the DEBRIS survey were used to place observational constraints on the radial location of the cold dust in this system. Resolving the cold dust is key to breaking the degeneracies inherent in SED modelling. We also use detection limits from unresolved images  at 12 and 18 $\micron$ from MICHELLE on Gemini, at 0.6 $\micron$ with ACS on HST, the unresolved Herschel SPIRE image at 250 $\micron$  and detailed SED modelling including all data from the literature to gain a complete picture of the disc. Modelling indicates that for a 2 component model of the system consisting of a hot and cold disc, the cold disc imaged with Herschel PACS lies between 15-70 AU, with a surface density profile $\Sigma \propto r^{-1.5}$ at an inclination of 55$^{\circ}$ from edge on, with the hot dust at 2 AU. SED fitting to observations from 5 $\micron$ to 1mm, including the IRS spectrum (\citealp{2006ApJS..166..351C})  suggest that the grain size distribution is consistent with that for a theoretical collisional cascade with a fixed maximum grain size of 1cm, a minimum grain size of 3 $\micron$ (0.3$\times D_{bl}$  for a composition with a silicate fraction of 20\%, a porosity of 20 \% with the rest of the grain composed of organic refractories. For the hot component the minimum grain size is 0.6 $\micron$ (0.1$\times D_{bl}$) and the silicate fraction was 60\%.  A 3 component model indicates that another possible configuration consist of an inner edge to the cold disc at 30 AU, a warm ring at 9 AU and hot dust at 2 AU. It is also possible to fit the observations of $\beta$ Leo using a single, very eccentric (e=0.92) planetesimal population, after \citet{2010MNRAS.402..657W}. This degeneracy implies that even a wealth of multi-wavelength data including the resolved images may not be enough to uniquely constrain the location of the dust when there are multiple populations and the edges of the belts have not been resolved.

$\beta$ Leo is a similar age to Fomalhaut and Vega, which also have hot dust within a few AU of the star. However, its cold disc is smaller and closer to the hot emission that either of these systems, and there may be an intermediate warm component. The most analogous disc among those resolved around A stars is Eta Tel, a 12 Myr star with a resolved disc from 21-26 AU and hot, unresolved dust at <4 AU (\citealp{2009AA...493..299S}), but in terms of the separation and ratio of the hot and cold components as shown in Figure \ref{A_stars}, $\beta$ Leo most resembles HD71155. 

We have also examined the population of A stars known to have both hot and cold dust discs, and we see two main 'types' of distributions between the hot and cold populations. First, discs such as  Vega and $\beta$ Leo which have a large (>20 AU) separation between the location of the hot and cold dust, and whose hot dust has a high 
fractional luminosity. The other type of disc, typified by Eta Tel and HR8799 show an almost flat or, for HR8799, declining line in Figure \ref{A_stars}, indicating that the hot dust has a similar or lower fractional luminosity than the cold component. In the case of HR8799 the hot and cold components are separated by a known planetary system  (\citealp{2008Sci...322.1348M}) which may inhibit the transfer of material from the outer to inner regions of the system.  However, the small number of stars in this sample and the possibility of unknown systematic issues from comparing dust detected through different methods limit the conclusions that can currently be made. 

\section{acknowledgments}

LJC  would like to acknowledge the support of an STFC studentship. The authors would like to thank James Di Francesco for his valuable comments which much improved the paper. The authors would also like to thank Andy Gibb for his help with the SCUBA-2 data reduction. Herschel is an ESA space observatory with science instruments provided by European-led Principal Investigator consortia and with important participation from NASA.

\bibliographystyle{mn2e} 
\bibliography{Beta_Leo_Refs}
 \label{lastpage} 

\end{document}